\documentclass[aps,prb,twocolumn,showpacs,10pt]{revtex4-1}

\usepackage{amsmath,amssymb}
\usepackage{graphicx}
\usepackage[caption=false]{subfig}
\usepackage{tikz}
\usetikzlibrary{arrows}
\usepackage{dsfont} %\mathds{1} is Identity

\newcommand\dg\dagger
\newcommand\R\rangle
\renewcommand\L\langle
\renewcommand\l\left
\renewcommand\r\right

\newcommand\er{\,\mathrm{e}}
\newcommand\dr{\,\mathrm{d}}
\newcommand\ir{\,\mathrm{i}}

\renewcommand\vector[1]{\mathbf{#1}}
\newcommand\av{\vector{a}}

\newcommand\ev{\vector{e}}

\newcommand\qv{\vector{q}}
\newcommand\Qv{\vector{Q}}
\newcommand\rv{\vector{r}}
\newcommand\Rv{\vector{R}}
\newcommand\uv{\vector{u}}

\DeclareMathOperator{\Pf}{Pf}

\newcommand{\nel}{n_{\mathrm{el}}}

\makeatletter
    \renewcommand\@make@capt@title[2]{%
     \@ifx@empty\float@link{\@firstofone}{\expandafter\href\expandafter{\float@link}}%
      {\textbf{#1}}\@caption@fignum@sep#2\quad}%
    \makeatother
\begin{document}

\title{Fractional quantum Hall states versus Wigner crystals in wide-quantum wells in the half-filled lowest and second Landau levels}

\author{N. Thiebaut$^1$, N. Regnault$^{2,3}$ and M. O. Goerbig$^1$}

\affiliation{$^1$Laboratoire de Physique des Solides, Universit\'e Paris-Sud, CNRS UMR 8502, F-91405 Orsay Cedex, France\\
$^2$Department of Physics, Princeton University, Princeton, New Jersey 08544, USA\\
$^3$
Laboratoire Pierre Aigrain, Ecole Normale Sup\'erieure-PSL Research
University, CNRS, Universit\'e Pierre et Marie Curie-Sorbonne Universit\'es,
Universit\'e Paris Diderot-Sorbonne Paris Cit\'e, 24 rue Lhomond, 75231
Paris Cedex 05, France}

\date{\today}

\begin{abstract}

We investigate numerically different phases that can occur at half filling in the lowest and the first excited Landau levels in wide-well twodimensional electron systems exposed to a perpendicular magnetic field. Within a twocomponent model that takes into account only the two lowest electronic subbands of the quantum well, we derive a phase diagram that compares favorably with an experimental one by Shabani \emph{et al.} [Phys. Rev. B \textbf{88}, 245413 (2013)]. In addition to the compressible composite-fermion Fermi liquid in narrow wells with a substantial subband gap and the incompressible twocomponent $(331)$ Halperin state, we identify in the lowest Landau level a rectangular Wigner crystal occupying the second subband. This crystal may be the origin of the experimentally observed insulating phase in the limit of wide wells and high electronic densities. 
In the second Landau level, the incompressible Pfaffian state, which occurs in narrow wells and large subband gaps, is also separated by an intermediate region from a large-well limit in which a similar rectangular Wigner crystal in the excited subband is the ground state, as for the lowest Landau level. However, the intermediate region is characterized by an incompressible state that consists of two four-flux Pfaffians in each of the components. 

\end{abstract}


\maketitle

%=====================================================================================
\section{Introduction\label{sec:Introduction}}

The occurance of correlated electronic liquids and crystalline phases in Landau levels (LLs) gives rise to a plethora of intriguing effects. Beyond the fractional quantum Hall effect,\cite{Tsui,Stormer1999} which is due to the formation of homogeneous incompressible liquids described with great success by trial wave functions,\cite{Laughlin1983b,Jain1989,Moore1991,Greiter1991} one finds Wigner crystals (WCs) that are responsible for the insulating phase at very low fillings of the lowest LL as well as more exotic bubble and stripe phases, \cite{Fogler1996,Fogler1997,MC} where the latter is responsible for an observed anisotropy in the longitudinal magnetoresistance.\cite{Lilly1999a,Lilly1999b,Du1999} These correlated phases arise due to the quantization of the kinetic energy of twodimensional electrons in a perpendicular magnetic field $B$ into highly degenerate LLs. The filling of these LLs is then characterized by the filling factor $\nu=\nel/n_B$, in terms of the electronic density $\nel$ and the density of states $n_B=eB/h$ in a LL. Unless an integer number of LLs is completely filled (integral values of $\nu$), the electronic interactions remain as the only relevant energy scale in the system, apart from the disorder potential which can at first be discarded in the description of the phases. These interactions are precisely at the origin of the above-mentioned correlated phases, e.g. the fractional quantum Hall states (FQHSs) which occur at certain ``magic'' fractions of the filling factor ($\nu=1/3, 1/5, 2/5, 5/2, \dots$ and many others). In higher LLs, the competition between electron solids and incompressible liquid phases is at the origin of the reentrant integer quantum Hall effect,\cite{Eisenstein2002}  where FQHSs at different filling factors are separated by insulating WC or a bubble crystal.\cite{Goerbig2003,Goerbig2004,Goerbig2004c,Goerbig2004d}

The nature of all these ground states and the competition between them are not only determined by the filling factor $\nu$, but also by the precise form of the effective in-plane interaction projected onto a given LL. The form of the latter is affected by the index $n$ of the partially filled LL, by screening, but also by the ``thickness'' of the quantum well, which hosts the twodimensional electron system (2DES). This situation occurs regularly because the highest mobility 2DES, which allow for the resolution of the more fragile FQHSs, are obtained in wide quantum wells with a typical width ranging from $40$ to $100$ nm. For experimentally relevant magnetic fields,
this corresponds to several (up to 20) magnetic lengths $\ell_B=\sqrt{\hbar/eB}\simeq 26$ nm$/\sqrt{B {\rm [T]}}$.
The well width can even lead to the formation of novel FQHSs, such as the ones observed at $\nu=1/2$ and $1/4$ in the lowest LL.\cite{Suen1994,Luhman2008,Shabani2009b,Shabani2013} The reminiscence with even-denominator states in bilayer quantum Hall systems \cite{Suen1992a} hints at a multi-component origin later confirmed theoretically both for symmetric\cite{Papic2009,Peterson2010a,Papic_Bilayer,Peterson2010b,Biddle2013} and asymmetric\cite{Scarola2010,Thiebaut2014} quantum wells in terms of a $(331)$ Halperin state \cite{Halperin1983} that competes with the compressible composite-fermion Fermi liquid (CFFL)\cite{Halperin1993} and, under certain circumstances,\cite{Peterson2010a,Papic_Bilayer,Scarola2010,Lu2010} with a Pfaffian.\cite{Moore1991,Greiter1991}
Furthermore, these liquid phases compete with an insulating phase in the limit of very large well widths.\cite{Shabani2013}

In the present paper, we investigate the competition between quantum liquids and WCs at half filling in wide quantum wells within a twocomponent model originally proposed in Ref. \onlinecite{Papic2009}. Within this model, only the two lowest electronic subbands, due to the well confinement potential, are taken into account whereas higher subbands are assumed to be unoccupied. The relevant parameters are the width of the quantum well and the subband gap that separates the lowest from the first excited electronic subband. Whereas these two parameters are in principle not independent of each other, they can be tuned effectively via the magnetic field, which determines at the same time the magnetic length $\ell_B$ and the Coulomb energy $e^2/4\pi\epsilon\epsilon_0\ell_B$ that are the natural scales in the system. We investigate various relevant FQHSs with the help of variational Monte-Carlo calculations (VMC) \cite{vonNeumann1951} in the spherical geometry whereas the energies of the electron-solid phases are calculated within the Hartree-Fock approximation (in the planar geometry). In spite of the difference in these approaches and in the geometries, the different energies can be compared directly in the thermodynamic limit, which we extract from the usual $1/N$ scaling of the VMC results. The obtained phase diagrams allow us to identify the different phases, which are responsible for the compressible, incompressible and insulating phases identified in magneto-transport measurements at half filling in the lowest LL.\cite{Shabani2013} Most saliently, we obtain an insulating rectangular WC in the excited subband as the ground state in wide quantum wells with low subband gap, both in the lowest and the first excited LL. This phase indicates a tendency to form stripes at half filling and adds to the numerous different types of WCs observed in wide quantum wells at other filling factors.\cite{Hatke2014,Hatke2014b} Whereas the intrinsic twocomponent state responsible for the fractional quantum Hall effect in the lowest LL is a $(331)$ Halperin state, in agreement with previous studies,\cite{Papic2009,Peterson2010a,Papic_Bilayer,Peterson2010b} our VMC favor an unusual FQHS in the second (or first excited) LL that consists of two four-flux Pfaffians in each subband component. This state separates the usual monocomponent Pfaffian\cite{Moore1991,Greiter1991} in narrow quantum wells from the above-mentioned WC in wide quantum wells with low subband gaps. 

Our paper is organized as follows. We present the twocomponent model for wide quantum wells in Sec. \ref{sec:Model}. This model is used in the VMC analysis of the various quantum liquid phases in Sec. \ref{sec:Liquids}, while Sec. \ref{sec:WCs} is devoted to the energy calculations of the different electronic crystals within the Hartree-Fock approximation. Because Secs. \ref{sec:Liquids} and \ref{sec:WCs} are concerned solely with liquid and 
crystalline phases, respectively, they may be considered as intermediate steps for our main results for the overall phase diagrams in Secs. \ref{sec:PD} and \ref{sec:LL1}. They may therefore be skipped by readers interested by the resulting phase diagrams. However, they contain technical details and additional information that we consider sufficiently important not to relegate them to an appendix. 
We discuss the theoretical phase diagram for the half-filled lowest LL in Sec. \ref{sec:PD} in comparison with experimental findings, and briefly discuss the phase diagram of the half-filled second LL in Sec. \ref{sec:LL1}. 

%=====================================================================================
\section{Wide quantum well model\label{sec:Model}}

Most experiments on quantum Hall effects are performed in Al$_x$Ga$_{1-x}$As/GaAs quantum wells that are more than 40 nm wide.~\cite{Suen1992a,Suen1992b,Suen1994,Manoharan1996,Shayegan1996,Shabani2009a,Shabani2009b} The lowest subbands of the latter are very well approximated by the lowest subbands of an infinite square quantum well~\cite{Shik}. \footnote{ The depth of the well is much larger than the subband gap for typical dopings ($x\simeq 30$\%).}  For this reason we model the confinement potential by an infinite square quantum well of width $w$,
in which the energy of the $\alpha$-th subband is $E_\alpha=\alpha^2 \hbar^2\pi^2/(2mw^2)$, where $m$ is the effective mass of the electrons. The corresponding eigenfunctions are given by
\begin{equation}\label{eq:SQWEigenfunctions}
\varphi_{\alpha}(\zeta) = \sqrt{\frac{2}{w}}\times \l\{ \begin{array}{cc}
\cos \left( \frac{\alpha \pi \zeta }{w} \right) &  \text{if $\alpha$ is odd}  \\ & \\
\sin \left( \frac{\alpha \pi \zeta }{w} \right) & \text{if $\alpha$ is even,}  \\
\end{array} \r. 
\end{equation}
where the $\zeta$ variable is restricted to $[-\tfrac{w}{2},\tfrac{w}{2}]$.
Although the Fermi energy usually lies below the first excited subband ($\alpha=2$) for non-interacting electrons, the $N$-particle ground state may be characterized by a non-zero occupation of higher subbands due to electron-electron interactions. Indeed, the presence of nodes in the excited subbands' eigenfunctions lowers the interaction energy, and in very wide wells this energy gain may overcome the gap between the first and the second subband
\begin{equation}
\Delta = \frac{3\pi^2}{2}\, \frac{\hbar^2}{m w^2},
\end{equation}
in which case the excited subbands are populated. On the other hand, the subband energy quickly increases with the subband index $\alpha$ ($E_\alpha \propto \alpha^2$), such that the ground state is unlikely to involve subbands beyond the first excited one, at least for realistic well widths. In our model, we will only retain the lowest two subbands and justify this assumption \emph{a posteriori} for the widest wells considered. Hence we consider in the following a twocomponent system the two components of which represent the first ($\alpha=1$) and the second ($\alpha=2$) subband. Similarly to bilayer quantum Hall systems, this twocomponent system may conveniently be described in terms of a pseudospin 1/2, where the two eigenvalues of the pseudospin projectors correspond to the two subbands.\cite{Papic2009}
In this analogy, the energy difference $\Delta$ between two subbands plays the role of an effective Zeeman effect that favors the occupation of the lowest subband. Whereas one could in principle also take into account the true spin of the electrons, within a four-component description, we rely on the fact that strong exchange interactions favor a full spin polarization at half-filling that is eventually oriented by the Zeeman field. The spin is thus effectively frozen out and omitted in the remainder of this paper. 

The three-dimensional Coulomb interaction
\begin{equation}\label{Ham:int}
 V\left(\rv, |\zeta-\zeta '|\right)=\frac{e^2}{4\pi\epsilon_0 \epsilon}\frac{1}{\sqrt{\rv^2 +|\zeta -\zeta '|^2}},
\end{equation}
where $\rv=(x,y)$ is the position in the plane, gives rise to an effective twodimensional Coulomb interaction
\begin{multline}\label{eq:potential}
V^{\alpha_1\dots \alpha_4}(r)=  \frac{e^2}{4\pi\epsilon_0 \epsilon} \int_{-\tfrac{w}{2}}^{\tfrac{w}{2}}  \int_{-\tfrac{w}{2}}^{\tfrac{w}{2}}  \\ \times \dfrac{\varphi_{\alpha_1}(\zeta)\varphi_{\alpha_2}(\zeta')\varphi_{\alpha_3}(\zeta)\varphi_{\alpha_4}(\zeta')}{\sqrt{r^2+(\zeta-\zeta')^2}} d\zeta d\zeta',
\end{multline}
that depends on the four subband components $\alpha_1$, $\dots$, $\alpha_4$ through the integration over the associated wave functions~(\ref{eq:SQWEigenfunctions}). In the above expressions, $\epsilon_0$ is the permittivity of the vacuum and $\epsilon$ is the static dielectric constant of the host material. In the remainder of this paper, the energy unit is set to  $e^2/4\pi\epsilon_0 \epsilon \ell_B$.

When projected to the lowest LL, the full Hamiltonian only consists of a Coulomb interaction term and a subband term. It reads
\begin{equation}\label{eq:FullHamDiv}
H = \hat V + \frac{\Delta}{2}\left(\hat{n}_{\alpha=2}-\hat{n}_{\alpha=1}\right)
%\sum_{\alpha} E_\alpha \hat{n}_\alpha
\end{equation}
where the interaction operator is
\begin{multline}\label{eq:IntOp}
\hat V = \frac12 \sum_{\l\{\alpha_i\r\}} \int \psi_{\alpha_1}^\dg(\rv) \psi_{\alpha_3}(\rv) \\ \times V^{\alpha_1\dots \alpha_4}(\l| \rv-\rv ' \r|)\psi_{\alpha_2}^\dg(\rv ')  \psi_{\alpha_4}(\rv ')  d^2r  d^2 r'
\end{multline}
in terms of the effective potential~(\ref{eq:potential}) and the twodimensional electronic field operators
$\psi_{\alpha}(\rv)$ in the lowest LL and subband $\alpha$. Note that the usual normal ordering is taken into account by a proper regularization detailed in the appendix.
Furthermore, $\hat{n}_\alpha=\int d^2r\psi_{\alpha}^{\dagger}(\rv)\psi(\rv)_{\alpha}$ is the number operator that counts the electrons in the subband $\alpha$.

For each FQHS, we compute the average value of the interaction energy~(\ref{eq:FullHamDiv}) for a given number of electrons $N$ by means of the variational Monte-Carlo method that is described in the appendix. The corresponding energy diverges in the thermodynamic limit, so that we choose to substract a regularization energy and divide by $N$ in order to obtain a quantity that is finite even when extrapolated to $N\rightarrow\infty$; we call this quantity the \emph{cohesive energy}.

With regard to the subband degree of freedom, the trial states under scrutiny can be divided into two groups~: monocomponent states, in which all electrons are in the same subband, and truly twocomponent states for which electrons necessarily occupy both components.\footnote{A true twocomponent state cannot be transformed via a global rotation into a one-component state. In principle one should also consider monocomponent states made out of linear combinations of the subbands, but the density profiles of such states are not symmetric around the center of the well and we expect the lowest energy state to be symmetric in a symmetric well.} 
Naturally, one expects monocomponent states (in $\alpha=1$) in the narrow-well limit, where the subband gap 
$\Delta$ is much larger than the interaction energy. The latter intervenes in the energy gain when populating
the subband $\alpha=2$ -- because of the node in the wave function $\varphi_{\alpha=2}(\zeta)$, the Coulomb
repulsion is reduced in $\alpha=2$. From a theoretical point of view, one may thus characterize a wide well as
one with a width such that $e^2/4\pi \epsilon_0\epsilon l_B$ is on the order of or larger than $\Delta$. A
more detailed discussion of the resulting scaling relations can be found in Sec. \ref{sec:2compLiquids}, where
we analyze the phase diagram of the liquid phases. 

\begin{figure}[ht]
\centering
\begin{center}
\includegraphics[scale=0.6]{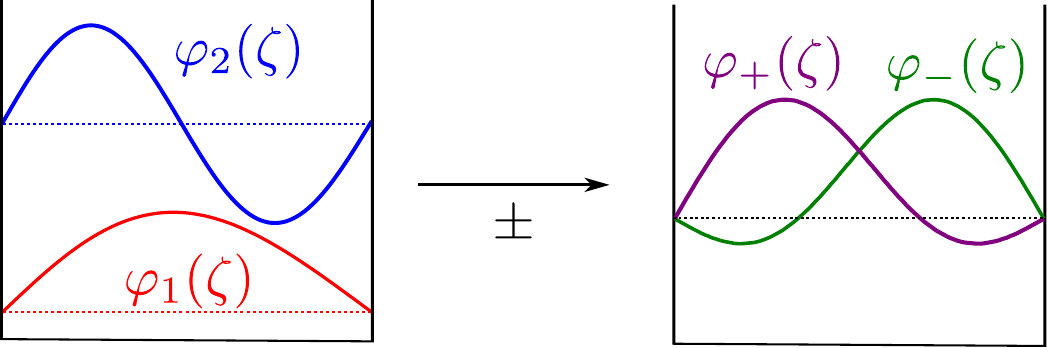}
\caption[]{(Color online) Lowest subband eigenfunctions (left) and effective bilayer wave functions (right). The latter are the $+$ and $-$ linear combinations of the former, \emph{i.e.} $\varphi_\pm=\left(\varphi_1\pm\varphi_2 \right)/\sqrt2$.}\label{fig:SubbandEigenfunctions}
\end{center}
\end{figure}

The notion of ``components'' does not necessarily refer to the subband index. Indeed, the components might refer to orthogonal combinations of the subband eigenstates.  
Here, we only consider the \emph{effective bilayer} configuration, in which the two components ``$+$'' and ``$-$'' are defined by  $\varphi_{\pm}=(\varphi_{\alpha=1}\pm\varphi_{\alpha=2})/\sqrt2$. The effective bilayer wavefunctions are shown in Fig.~\ref{fig:SubbandEigenfunctions}. This choice for the two components is natural since the density profile of a corresponding state has the $z\mapsto -z$ symmetry of the well. 
Our two-subband model thus accounts for the formation of an effective bilayer in wide wells due to the Coulomb
repulsion, as corroborated by self-consistent Hartree-Fock calculations.\cite{Suen1992b} However, we emphasize
that our model is not biased to the formation of effective-bilayer states or the particular choice of 
$\varphi_{\pm}$ for the twocomponent basis. Indeed, the latter choice is natural only in the limit of a vanishing effective subband gap, where the renormalization of the gap by interactions is taken into account.\cite{Papic2009} 

While monocomponent states may lie either in the lowest or in the first excited subband, all the twocomponent states that we will study are in the effective layer configuration, in which half of the electrons are in the $\varphi_+$ state and the other half is in the $\varphi_-$ state. This assumption is corroborated by previous studies by Ref.~\onlinecite{Papic2009} that showed that the pseudospin-1/2 symmetry-breaking terms in the interaction Hamiltonian (\ref{Ham:int}) favor an annihilation of the $x$-component of the subband pseudospin, i.e. its orientation in the $yz$-plane and a final orientation in the $z$-direction due to the subband gap. Furthermore, we have checked numerically in the present study that the lowest-energy monocomponent states have indeed a pseudospin polarization in the $z$-direction, whereas we have shown within exact-diagonalization studies via a variational parameter that the twocomponent states occupy equally the $\varphi_+$ and $\varphi_-$ states.\cite{Thiebaut2014}

%The energies of the liquids states are computed by means of the variational Monte-Carlo method, performed in the spherical geometry, while the energies of the solid phases are simply computed in the Hartree-Fock approximation, that is known to give accurate results in that case. Because 

%--------------------------
\section{Liquids in the lowest Landau level: composite-fermion Fermi liquid, $(331)$ Halperin and Pfaffian states\label{sec:Liquids}}
Before introducing the solid phases, we present our results for the relevant liquid states. As mentioned above, the monocomponent liquids, the CFFL and Pfaffian, can either be in the lowest or in the first excited subband. %It could in principle also be in a superposition of the two subbands, but the density profile of such a state does not have the $\zeta\mapsto -\zeta$ symmetry of the well, and so it is not likely to be favored. 

Concerning effective bilayer liquids, the $(331)$ Halperin state is not the only one that needs to be considered. 
Indeed, beyond the $(331)$ state we can also artificially superimpose monocomponent FQHSs, each with a filling factor $1/4$, with a spinor structure. By doing so one can construct the ``2CFFL'' state, made out of two uncorrelated CFFLs in each of the components, and the ``2Pf'' state built with two independent Pfaffian phases. These states are presented in detail in the following. 

\subsection{Monocomponent liquids}

With the help of the variational Monte-Carlo (VMC) procedure described in the appendix~\ref{sec:VMC}, we first compute the cohesive energies of the monocomponent liquids in the subband $\alpha$. The trial wave function for the CFFL$_{2n}$
\begin{eqnarray}\label{WF:CFFL}\nonumber
\psi_{\mathrm{CFFL}_{2n}}^{\alpha}\left(\l\{z_i^\alpha\r\}\right) &=& \mathcal{P}_{\mathrm{LLL}}\left[\det\left(\er^{-i(k_iz_j^\alpha+k_i^*z_j^{\alpha *})}\right) \right.\\
&&\times \left.\prod_{i<j}\left(z_i^\alpha-z_j^\alpha \right)^{2n}\right]
\end{eqnarray}
can be expressed in terms of the Slater determinant $\det[\exp(-i(k_iz_j^\alpha+k_i^*z_j^{\alpha *})]$, where 
$z_j^{\alpha}=x_j^{\alpha}+i y_j^{\alpha}$ is the complex coordinate of the $j$-th electron in the subband $\alpha$ and $k_j$ the associated (complex) wave vector. Since the wave functions in the determinant bear non-analytic components, in terms of the conjugate coordinate $z_j^{\alpha *}$, the overall wave function needs to be projected into the lowest LL by the projector $\mathcal{P}_{\mathrm{LLL}}$. Similarly, the Pfaffian reads
\begin{equation}\label{WF:Pfaff}
\psi_{\mathrm{Pf}_{2n}}^\alpha\left(\l\{z_i^\alpha\r\}\right)=\Pf\left(\frac{1}{z_i^\alpha -z_j^\alpha}\right) \prod_{i<j}\left(z_i^\alpha-z_j^\alpha \right)^{2n},
\end{equation}
where $\Pf$ denotes the Pfaffian of the matrix $M_{ij}=1/(z_i^{\alpha}-z_j^{\alpha})$. In both wave functions (\ref{WF:CFFL}) and (\ref{WF:Pfaff}), we have omitted the obiquous Gaussian factor $\exp(-\sum_i^{N}|z_i^{\alpha}|^2/4)$ for the lowest LL in order to simplify the notations. Moreover, they are both associated with the filling factor $\nu=1/2n$.

\begin{figure}[ht]
\begin{center}
\includegraphics[scale=1]{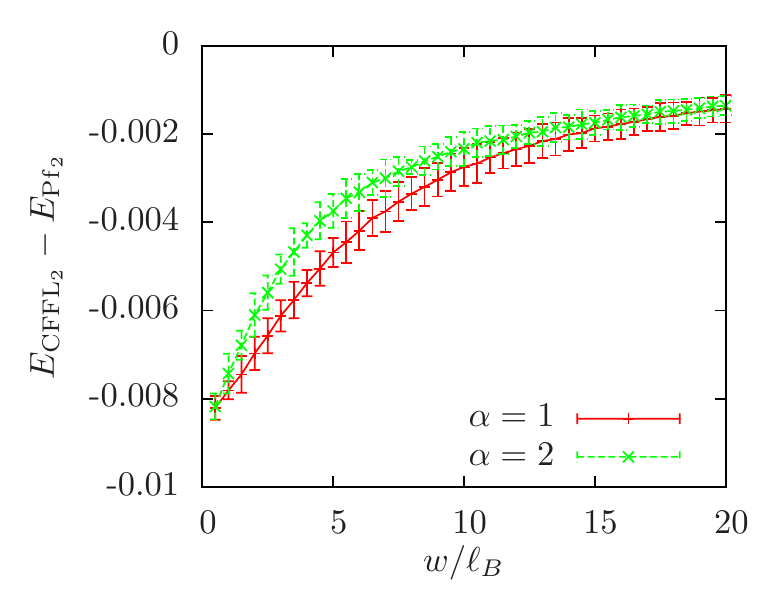}
\caption{(Color online) Cohesive-energy difference of the CFFL$_2$ and the Pf$_2$ in the lowest subband ($\alpha=1$) and in the first excited subband ($\alpha=2$) as a function of the well width $w/\ell_B$. The CFFL$_2$ is always lower in energy.}
\label{fig:PfVsCFFL}
\end{center}
\end{figure}

For monocomponent states, we consider $n=1$. The difference in cohesive energy of the CFFL$_2$ and the Pfaffian is plotted in Fig.~\ref{fig:PfVsCFFL} 
as a function of the well width for both subbands. The Pf$_2$ state in the lowest subband ($\alpha=1$)
is known to have a higher energy than the CFFL$_2$ for $w=0$ in the lowest LL from previous studies.\cite{Park1998,Biddle2013} Figure~\ref{fig:PfVsCFFL} shows that the CFFL$_2$ remains lower in energy than the Pf$_2$ at least up to $w/\ell_B=20$, although the energies get closer with increasing width. This piece of information is noteworthy since in wide quantum wells, the softening of the Coulomb interaction leads to a reduced ratio $V_1/V_3$ of the first two odd pseudopotentials compared to the pure Coulomb case, and this reduction is generally expected to be in favor of the Pfaffian state.\cite{Morf1998} As mentioned above, the error bars for the energy difference are always larger than the uncertainty in the $1/N=0$ extraction shown in Fig.~\ref{fig:ExempleInterpolation}. 

\begin{figure}[ht]
\begin{center}
\includegraphics[scale=1]{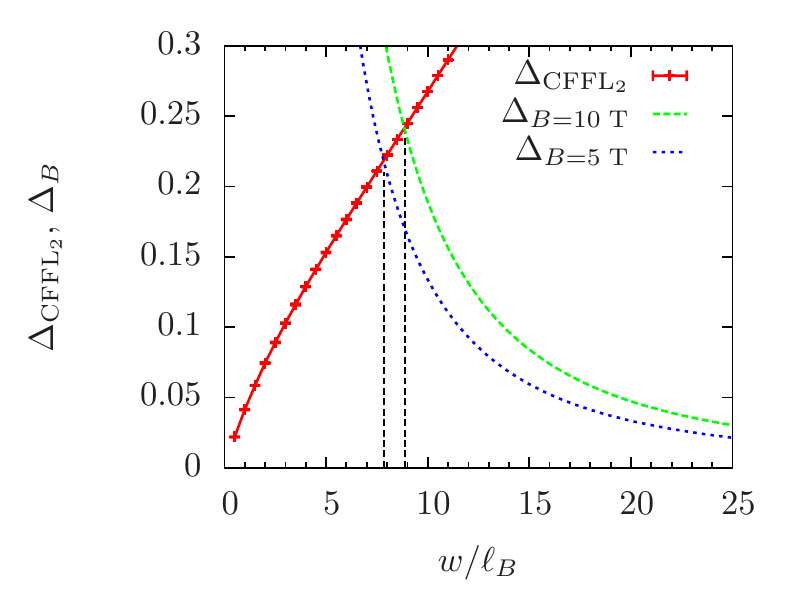}
\caption{(Color online) Energy difference $\Delta_{\mathrm{CFFL}_2}=E_{\mathrm{CFFL}_2}^{\alpha=1}-E_{\mathrm{CFFL}_2}^{\alpha=2}$
between the CFFL$_2$ in the lowest and that in the first excited subband (red line). We have plotted the subband gap in the Coulomb energy unit $\Delta_B=\Delta/(e^2/4\pi\epsilon\epsilon_0\ell_B)$ as a function of the width of the well $w/\ell_B$ for $B=10$ T (green dashed line) and $B=5$ T (blue dotted line) for comparison.} %The transition occurs at smaller widths for small magnetic fields.}
\label{fig:CFFLLowestVsExcited}
\end{center}
\end{figure}

In the first excited subband ($\alpha=2$), although the energies of the Pf$_2$ and CFFL$_2$ are different from those computed in the lowest subband, their relative  difference is almost the same (Fig.~\ref{fig:PfVsCFFL}), such that also in the first excited subband the Pf$_2$ cannot be lower in energy than the CFFL$_2$. 
One notices that the energies in the excited subband are systematically lower than those in the lowest one 
($\alpha=1$). This reflects the abovementioned node in the wavefunctions in the $z$-direction, which reduces the 
Coulomb repulsion in $\alpha=2$ as compared to $\alpha=1$, and results in the tendency to spontaneously form bilayer states.\cite{Suen1992b} However, we emphasize that one
cannot compare states that belong to different subbands directly, since one has to specify the subband gap to do so. In Fig.~\ref{fig:CFFLLowestVsExcited}, we have plotted the energy difference $\Delta_{\mathrm{CFFL}_2}=E_{\mathrm{CFFL}_2}^{\alpha=1}-E_{\mathrm{CFFL}_2}^{\alpha=2}$
between the CFFL$_2$ in the lowest and that in the first excited subband, neglecting the subband gap. The latter is plotted as an independent quantity for two values of the magnetic field ($B=5$ T and $B=10$ T). Indeed, one notices that in the absence of $\Delta$, it is always favorable to form a CFFL in the first excited subband because of the lower interaction energy. If we now take into account the subband gap, the energy difference defines a gap value $\Delta_\mathrm{CFFL}$ below which the CFFL prefers the first excited subband, in spite of the cost in interaction energy. In units of the typical Coulomb energy $e^2/4\pi\epsilon\epsilon_0\ell_B$ the subband gap reads $\Delta\simeq 5.93\sqrt{B}/(w/\ell_B)^2$, such that the transition from the lowest subband to the first excited subband CFFL occurs at a width $w/\ell_B$ that depends on $B$. For instance for $B=5$ T this transition occurs at $w/\ell_B\simeq 8$ while it happens at $w/\ell_B\simeq 9$ for $B=10$ T (see Fig.~\ref{fig:CFFLLowestVsExcited}).

\subsection{Twocomponent liquids}

In addition to the above-mentioned monocomponent trial states, we investigate here the energies of two types of twocomponent states. The first one consists of states with no correlations between the components, namely the ``uncorrelated CFFLs'' (2CFFL$_4$), which consist of two CFFLs at $\nu_\pm=1/4$ in each component, and the ``two uncorrelated Pfaffians'' (2Pf$_4$) composed of four-flux composite fermions.
In terms of the complex positions $z_j^{\pm}$ of the $j$-th electron in the effective-bilayer state $\pm$, the 2CFFL$_4$ wave function reads
\begin{equation}
\begin{aligned}\label{WF:1}
&\psi_{\mathrm{2CFFL}_4}\left(z_1^+,\dots , z_{N/2}^+,z_1^-, \dots ,z_{N/2}^-\right)\\
&=  \psi_{\mathrm{CFFL}_4}\left(z_1^+,\dots , z_{N/2}^+\right)\psi_{\mathrm{CFFL}_4}\left(z_1^-, \dots ,z_{N/2}^-\right) \\
\end{aligned} 	
\end{equation}
where  $\psi_{\mathrm{CFFL}_4}$, given by Eq.~(\ref{WF:CFFL}) for $n=2$,
is the four-flux CFFL wave function. The 2Pf$_4$ wave function is defined analogously as a product of two four-flux Pfaffian states
\begin{equation}\label{WF:3}
\psi_{2\mathrm{Pf}_4}^\pm\left(\l\{z_i^\pm\r\}\right)=\psi_{\mathrm{Pf}_4}\left(z_1^+,\dots , z_{N/2}^+\right)\psi_{\mathrm{Pf}_4}\left(z_1^-, \dots ,z_{N/2}^-\right),
\end{equation}
where $\psi_{\mathrm{Pf}_4}$ is defined by Eq.~(\ref{WF:Pfaff}).
Finally we investigate the $(331)$ Halperin state~\cite{Halperin1983}
\begin{equation}
\begin{aligned}\label{WF:4}
&\psi_{(331)} \left(z_1^+,\dots , z_{N/2}^+,z_1^-, \dots ,z_{N/2}^-\right)\\
&=  \prod_{i<j}\left(z_i^+-z_j^+\right)^3  \prod_{k<l}\left(z_k^--z_l^-\right)^3 \prod_{m,n}\left(z_m^+-z_n^-\right)
\end{aligned}
\end{equation}
in which the electrons in both components are now cross-correlated by the last term. Similarly to the monocomponent wave functions (\ref{WF:CFFL}) and (\ref{WF:Pfaff}), we have omitted in all trial wave functions (\ref{WF:1}), (\ref{WF:3}) and (\ref{WF:4}) the Gaussian factor 
$\exp[-(\sum_i^{N_+}|z_i^{+}|^2 + \sum_i^{N_-}|z_i^{-}|^2)/4]$ for the lowest LL. Notice that one can also 
construct another Halperin state, $(113)$ with inverted exponents, at $\nu=1/2$. This state is however unstable
and subject to phase separation\cite{DeGail} so that we do not consider it here. 

\begin{figure}[ht]
\begin{center}
\includegraphics[scale=1]{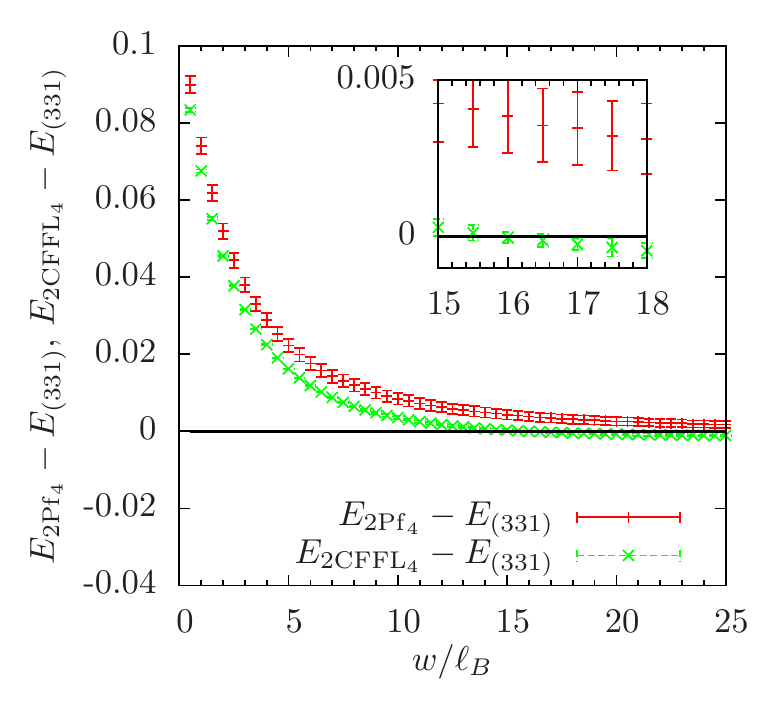}
\caption{(Color online) Cohesive energies of the effective-layer liquid states. The reference chosen here is the cohesive energy of the $(331)$ state. The inset is a zoom around the transition between the $(331)$ state and the 2CFFL$_4$. Notice that the error bars prevent the precise location of the transition (between $w/\ell_B\simeq 15$ and $w/\ell_B\simeq 17$).}
\label{fig:EffectiveBilayerLiquids}
\end{center}
\end{figure}

The cohesive-energy differences of those states are compared in Fig.~\ref{fig:EffectiveBilayerLiquids}. We have chosen the cohesive energy of the $(331)$ state as the reference. The latter has the lowest energy for $w/\ell_B \le 15$ while the 2CFFL$_4$ is favored in very wide wells ($w/\ell_B \ge 17$). 
The precise value of the width at which the transition occurs cannot be extracted from our numerical results due to uncertainties associated with the VMC and the polynomial extrapolation (see inset of Fig.~\ref{fig:EffectiveBilayerLiquids}).
This result can be understood quite intuitively: in a wide quantum well, correlations between the different components become less relevant, similarly to bilayer systems in the large-distance limit. One therefore retrieves two independant onecomponent systems with $\nu_\pm=1/4$, where the ground state is the 2CFFL$_4$.
\cite{Eisenstein1992,Toke2005} Just as in the monocomponent case the 2Pf$_4$ state is systematically higher in energy than the 2CFFL$_4$ and the (331) state, hence according to our calculations no Pfaffian phase can be stabilized in the lowest LL in a wide quantum well.

\subsection{Phase diagram of mono- and twocomponent liquid states}
\label{sec:2compLiquids}

\begin{figure}[ht]
\begin{center}
\includegraphics[scale=1]{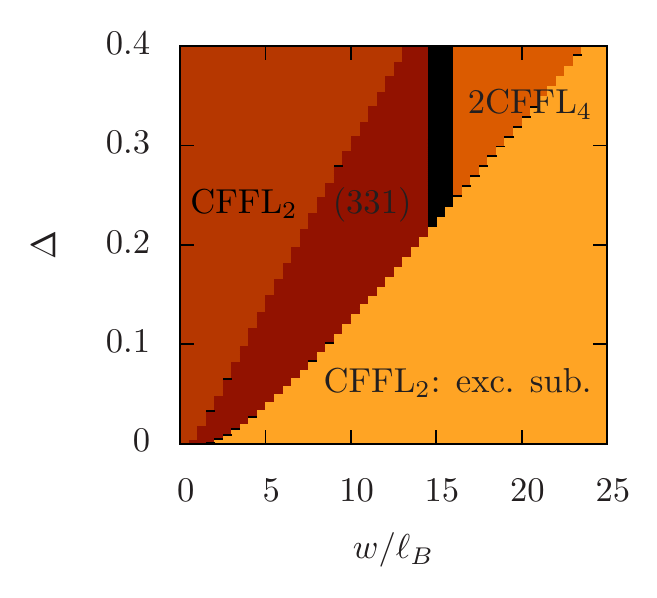}
\caption{(Color online) Phase diagram for liquid states. Black indicate regions where the uncertainties of the lowest-energy state overlaps with the uncertainty of another state, such that we cannot conclude. No Pfaffian state is stabilized here. Note that the black region of undeterminacy that separates the 2CFFL$_4$ and $(331)$ phases is rather large since the two phases have close energies for $w/\ell_B\simeq 15$, and they have the same subband energy per particle $E_{\mathrm{subband}}^{\mathrm{bil.}}=\Delta/2$. }
\label{fig:LiquidPhaseDiagramWidthDelta}
\end{center}
\end{figure}

To be relevant for realistic systems, the phase diagram of the liquid states should take into account the subband gap $\Delta$. Indeed, it adds a contribution \mbox{$E_{\mathrm{subband}}^{\mathrm{bil.}}=N\Delta/2$} to the effective bilayer states and a contribution $E_{\mathrm{subband}}^{\mathrm{exc.}}=N\Delta$ to the excited subband states. The phase  diagram is shown in Fig.~\ref{fig:LiquidPhaseDiagramWidthDelta} as a function of well width and subband gap. The subband gap is treated here as a parameter independant of the width for illustration.
As expected, large subband gaps favor the lowest subband CFFL$_2$ since the latter is the state with lowest energy in monocomponent systems. Conversely, when the well is large and for moderate subband gaps, it becomes favorable to populate the first excited subband due to the gain in interaction energy. In that regime (lower right part of the diagram), the CFFL$_2$ is also stabilized but now in the first excited subband.
These two regimes of monocomponent states are separated by a region in which the subband gap is on the same order of magnitude as the gain in interaction energy when populating the first excited subband, such that an effective subband gap that one can define on the mean-field level vanishes. In this approach, one finds for this region the scaling behavior for the correlation energy gain $\Delta_{\mathrm{2-comp}}$ of the twocomponent states compared to the lowest subband \cite{Papic2009}
\begin{equation}\label{Eq:scaling}
\frac{\Delta_{\mathrm{2-comp}}}{e^2/4\pi \epsilon_0\epsilon \ell_B} \propto \frac{w}{\ell_B}
\end{equation}
that is in good agreement with our results depicted in Fig. (\ref{fig:LiquidPhaseDiagramWidthDelta}).
In this region, one stabilizes the various twocomponent states of the effective bilayer, i.e. 
the 2CFFL$_4$ if the well is very large ($w/\ell_B\ge 17$) and the incompressible $(331)$ state for $w/\ell_B \le 15$.

\section{Wigner crystals in wide quantum wells in the lowest Landau level\label{sec:WCs}}

The FQHSs presented in the previous section are in close competition with differents types of Wigner crystals (WCs).\cite{Wigner1934,Lam1984} In contrast to the FQHSs, the latter are charge density waves of a particular type that can be described conveniently within the Hartree-Fock approximation of the interaction Hamiltonian.\cite{Fukuyama1979,Yoshioka1979,Yoshioka1983,Maki1983} They are relevant since, for instance, the triangular WC has a lower energy than all known FQHS at low densities~\cite{Lam1984,Esfarjani1990} $\nu\lesssim 1/6$, and it also shows up in the vicinity of FQHE magic filling factors through the reintrant integer quantum Hall effect.\cite{Santos1992,Li1991}

\subsection{Monocomponent Wigner crystals}

Let us first concentrate on monocomponent WCs in one of the subbands $\alpha$.
In WCs the electrons minimize the Coulomb interaction by being situated at the nodes of a lattice. The localized wave function of an electron around $\Rv$ with lowest kinetic energy is a coherent state
\begin{equation}
\Psi_\Rv(\rv)=\L\rv|\er^{\frac{R}{\sqrt{2}}b^\dg}|0,0\R =\sqrt{\frac{1}{2\pi}}\er^{\frac{\ir}{2}\left(y X - x Y\right)} \er^{-\frac{1}{4}\left( \rv - \Rv\right)^2}
\end{equation}
and the corresponding many-body state involves antisymetrization $\mathcal{A}$ of $N$ localized states, where $N$ is the number of electrons
\begin{equation}\label{eq:AntisymWC}
\Psi_{\mathrm{WC}}\left( \rv_1, \dots , \rv_N\right) = \mathcal{A}\left[ \Psi_{\Rv_1}(\rv_1) \dots \Psi_{\Rv_N}(\rv_N)\right].
\end{equation}
If we neglect the overlap of neighboring localized states, the probability density of the many-body state~(\ref{eq:AntisymWC}) is
\begin{equation}
\begin{aligned}
\rho(\rv_1) &= \int d^2r_2 \dots d^2r_N\l|\Psi_{\mathrm{WC}}\left( \rv_1, \dots , \rv_N\right)\r|^2 \\
\simeq\sum_{p\in\mathcal{\sigma}_N}& \int d^2r_2 \dots d^2r_N \l|\Psi_{\Rv_1}\left( \rv_{p(1)}\right)\r|^2\dots  \l|\Psi_{\Rv_N}\left( \rv_{p(N)}\right)\r|^2\\
& =  \frac{1}{2\pi}\sum_j \er^{-\frac{1}{2}\left( \rv - \Rv_j\right)^2},
\end{aligned}
\end{equation}
where $p$ is an element of the permutation group $\sigma_N$.
Neglecting the overlap between adjacent lattice sites is a reasonable assumption in the low-density limit, where
the average distance between electrons is much larger than the typical extent $\ell_B$ of the wavefunctions. In
the present case, however, we investigate WCs at $\nu=1/2$, where overlap corrections become relevant. In this case,
the square of the probability density in reciprocal space that is more convenient for our Hartree-Fock calculations reads
\begin{equation}
|\l\L \tilde{\rho}(\qv) \r\R|^2 =   \er^{-q^2/2}  |\l\L \overline{\rho}(\qv) \r\R|^2,
\end{equation}
whose explicit expression is derived in Appendix \ref{app:B} [Eq. (\ref{eq:density_corrections})].

In terms of the modulus square of the projected density $|\l\L \overline{\rho}(\qv) \r\R|^2$, which plays
the role of the crystalline order parameter, the energy of WCs in the Hartree-Fock approximation reads~\cite{Yoshioka}
\begin{equation}
E_\mathrm{HF} = \frac{1}{2A}\sum_\qv \tilde v_{\mathrm{HF}}(q)  \;\l| \l\L \overline{\rho}(\qv)\r\R\r|^2
\end{equation}
where $A$ is the area of the system,
%\begin{equation}
%\l\L \overline{\rho}(\qv) \r\R =\er^{\frac{q^2}{4}} \tilde{\rho}(\qv)  = N \er^{-\frac{q^2}{4}}\sum_{i}\delta\left(\qv-\Qv_i\right)
%\end{equation}
%is the lowest LL-projected density, with the vectors $\Qv_i$ of the WC reciprocal lattice, 
and $\tilde v_{\mathrm{HF}}(q)=\tilde v_\mathrm{H}(q)-\tilde v_\mathrm{F}(q)$ is the Hartree-Fock potential. 
The direct (Hartree) term 
\begin{equation}\label{eq:Hartree}
\tilde v_\mathrm{H}(q)=\tilde{v}(q)\exp(-q^2/2)
\end{equation}
is the lowest-LL-projected Fourier transform of the potential $V(r)$, and the exchange (Fock) term
\begin{equation}
\tilde{v}_\mathrm{F}(q) =2\pi \ell_B \;v_\mathrm{H}\left(q\ell_B^2\right)
\end{equation}
is proportional to the (inverse) Fourier transform of $\tilde{v}_H$ for a rotationally invariant interaction potential. Notice that the formalism is readily generalized to higher LLs with index $n$ ($n=0$ being the lowest LL) via the replacement of 
the Gaussian $\exp(-q^2/2)$ in Eq. (\ref{eq:Hartree}) by the complete form factor $\exp(-q^2/2)[L_n(q^2/2)]^2$, in terms of the $n$-th Laguerre polynomial. %[N.B. Remember that due to the counting that starts with $n=0$ for the lowest LL, the second (or first excited LL) has $n=1$.]

The density $\nel=N/A$ is related to the filling factor $\nu$ by $\nel=\nu/2\pi\ell_B^2$, so the energy per particle finally reads
\begin{equation}\label{eq:HFEnergy}
\frac{E_\mathrm{HF}}{N} = \frac{\nu}{4\pi}\sum_\qv \tilde v_{\mathrm{HF}}(q)  \;\er^{-\frac{q^2}{2}}.
\end{equation}
In the lowest subband the relevant potential is $V^{1111}$ [see Eq.~(\ref{eq:potential})], it is $V^{2222}$ in the first excited subband, and in the effective bilayer configuration we have to specify both the intra-layer interaction and the inter-layer one.
As mentionned in Sec.~\ref{sec:Liquids} (and discussed in the appendix) the energy of Eq.~(\ref{eq:HFEnergy}) needs to be regularized to obtain the cohesive energy $E_\mathrm{coh.}$. We have used the same procedure as for the one for the liquid state (see the appendix).
%Similarly to the previous section we substract the regularization energy~(\ref{eq:RegularizationEnergy}) per particle to the energy~(\ref{eq:HFEnergy}) to obtain the cohesive energy $E_\mathrm{coh.}$, which allows one to eventually compare the energies of the different liquid and solid phases. 
The $q=0$ term of the sum in the energy~(\ref{eq:HFEnergy}) is (partly) canceled by the regularization energy, and the rest of the sum converges rapidly thanks to the Gaussian factors $\exp(-q^2/2)$, such that only a hundreds of terms are enough to reach machine precision.

\begin{figure}
\centering
\includegraphics[scale=1]{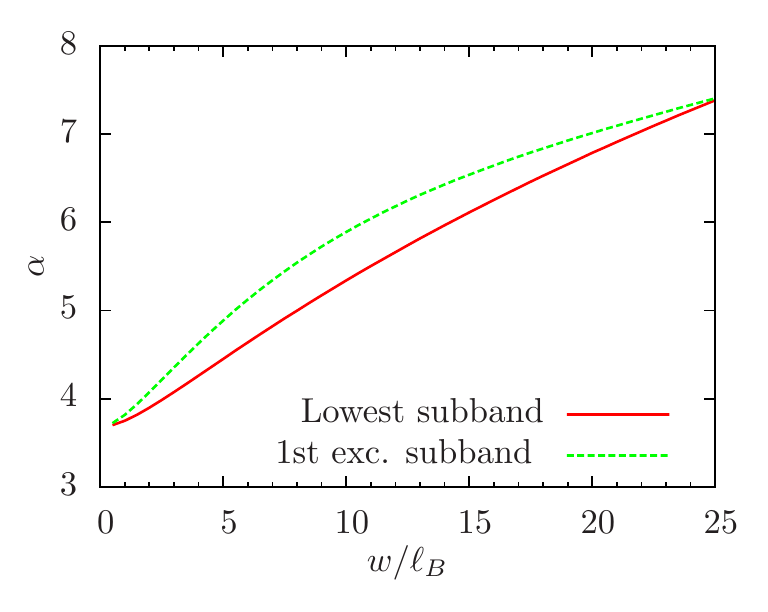}
\caption{(Color online) Anisotropy parameter $\alpha$ of the rectangular Wigner crystal with lowest energy, in the lowest and first excited subband, as a function of the well width $w$. The lattice becomes more anisotropic with increasing width.} 
\label{fig:AnisotropyParameters}
\end{figure}

We now have to determine the lattice that minimizes the energy of the crystal for both of the lowest subbands. We consider a general (monoclinic) twodimensional reciprocal lattice and minimize the cohesive energy with respect to the vectors that define the elementary cell, while keeping the area of the unit cell fixed to $1/\nel$. In doing so we  find that the minimum is obtained for a rectangular lattice, in both subbands and for all well widths. The basis vectors of this rectangular lattice read 
\begin{equation}
\av_1=\frac{1}{\sqrt{\nel}}(\sqrt{\alpha},0)\qquad \text{and}\qquad \av_2=\frac{1}{\sqrt{\nel}}(0,1/\sqrt{\alpha}),
\end{equation}
 where we have introduced the \emph{anisotropy parameter} $\alpha>1$ such that $\|\av_1\|=\alpha \|\av_2\|$. 
With this choice, the lattice sites are approached in the $y$-direction, whereas the lattice spacing is increased
in the $x$-direction.
The anisotropy parameters of the crystal with lowest cohesive energies are shown in Fig.~\ref{fig:AnisotropyParameters}. 
The Wigner crystal with lowest energy appears to have a non-negligible anisotropy even in thin wells, and this anisotropy grows with increasing well width. Anisotropic WCs are peculiar to half-filled LLs, and were interpreted as cousins of stripe phases in an earlier work (in the third LL).\cite{Ettouhami2006} Interestingly, we find even in the lowest LL (if we artificially discard the quantum-liquid phases) a tendency of the WC to form stripe-like phases. Notice that we have considered also the usual triangular WC, but it is higher in energy than the rectangular one.

\subsection{Twocomponent Wigner crystals}
\begin{figure}
\begin{center}
\includegraphics[width=8cm]{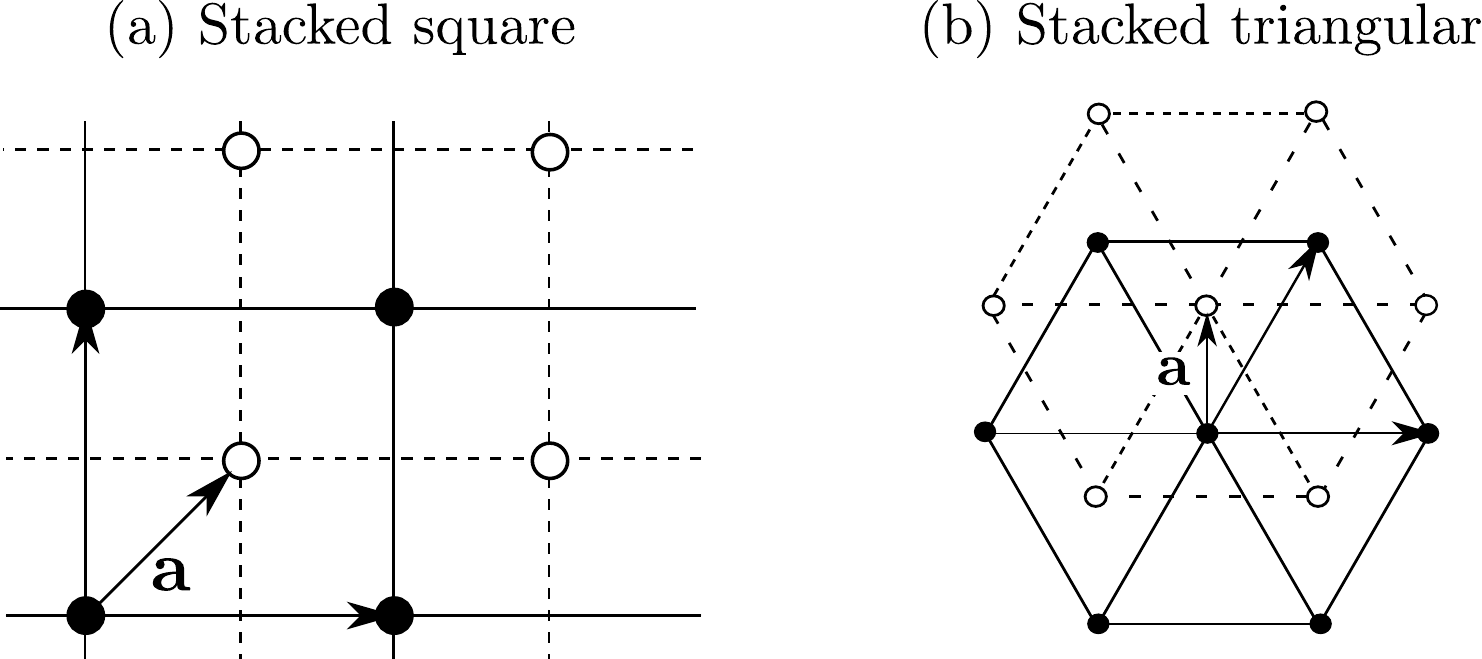}
\caption{The stacked square and stacked triangular lattices. The lattice of the $-$ layer (empty circles) is shifted with respect to the one of the $+$ layer (filled circles) by a displacement vector $\av$.} 
\label{fig:StackedLattices}
\end{center}
\end{figure}

As for the liquid phases, in addition to the monocomponent states that are either entirely contained in the lowest or in the first excited subband, we also have to consider effective bilayer WCs that consist of a WC in each of the effective layers. In order to minimize the interlayer repulsion the WC of one layer is positioned at the interstitials of that in the other layer. 
Here, we only present two types of lattices for the two component WC, the stacked square and stacked triangular lattices that are depicted on Fig.~\ref{fig:StackedLattices}. In those WCs half of the electrons are in the $+$ effective layer, while the other half is in the $-$ layer and localized around the sites of a lattice that is shifted by a displacement vector $\av$ with respect to the one of the $+$ layer. We have also considered more general stacked WCs such as the stacked rectangular one, but they happen to be higher in energy contrarily to the monocomponent crystals.

The intra- and inter-layer Hartree potentials are given by 
\begin{equation}
\tilde v_\mathrm{H}^{(\mathrm{A/E})}(q) = \tilde v^{(\mathrm{A/E})} (q) \er^{-\frac{q^2}{2}}
\end{equation}
where $\tilde v^{(\mathrm{A})}$ and $\tilde v^{(\mathrm{E})}$ are the Fourier transforms of the intra-layer effective potential~(\ref{eq:IntraLayerEffPot}) and inter-layer effective potential~(\ref{eq:InterLayerEffPot}), respectively. Careful handling of the Hartree-Fock approximation shows that the corresponding Fock terms are $\tilde v_\mathrm{F}^{(\mathrm{A})}(q) = 2\pi \ell_B \, v_\mathrm{H}^{(\mathrm{A})}(q\ell_B^2)$ and $\tilde v_\mathrm{F}^{(\mathrm{E})}(q) = 2\pi \ell_B \, v_\mathrm{H}^{(\mathrm{X})}(q\ell_B^2)$, such that the inter-layer Fock potential $\tilde v_\mathrm{F}^{(\mathrm{E})}$ is related to the cross term~(\ref{eq:CrossLayerEffPot}) and not to the inter-layer potential~(\ref{eq:InterLayerEffPot}).

In terms of those potentials the Hartree-Fock energy per particle of a stacked WC reads
\begin{equation}
\label{eq:FullBilayerHamiltonian}
\frac{E_\mathrm{stacked}}{N} = \frac{\nu}{2\pi}\sum_\qv \left[  \tilde v_\mathrm{HF}^{(\mathrm{A})}(q) 
 +\tilde  v_\mathrm{HF}^{(\mathrm{E})}(q)\cos(\qv\cdot\av) \right]\l| \l\L \overline{\rho}(\qv)\r\R\r| ^2 ,
%\\&+2\left[ v_H^{+--+}(q) - v_\mathrm{F}^{+-+-}(q)\right] \l|\L\overline{\rho}_{+-}(\qv)\R\r|^2\Big\}.
\end{equation}
where the reciprocal lattice that defines $\l\L \overline{\rho}(\qv)\r\R$ accomodates only half of the electrons, and $\vector{a}$ is the relative displacement vector of the stacked lattices (see Fig.~\ref{fig:StackedLattices}).

\begin{figure}
\includegraphics[scale=1]{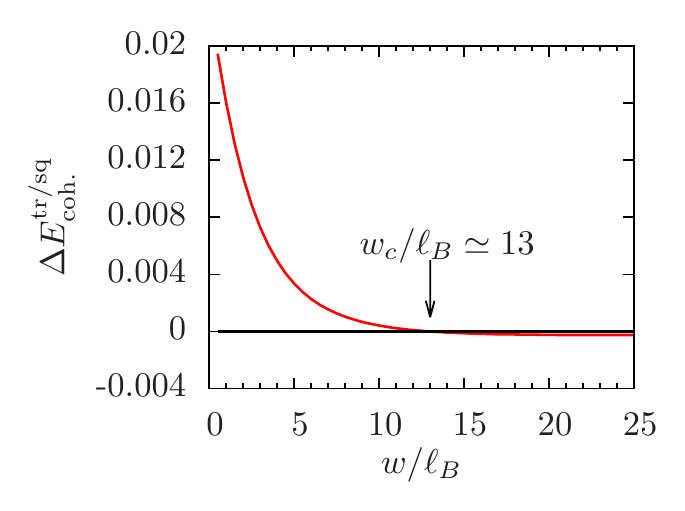}
\caption{(Color online) Cohesive energy difference $\Delta^{\mathrm{tr/sq}}_\mathrm{coh.}$ of the stacked triangular and stacked square Wigner crystals as a function of the well width $w$. The stacked square crystal has lower energy for $w/\ell_B\le 13$, while the stacked square lattice is favored in larger wells.} 
\label{fig:StackedEnergies}
\end{figure}

The energies of the two crystals are compared in Fig.~\ref{fig:StackedEnergies}, where we plot the difference in cohesive energy between the stacked triangular and the stacked square crystals.
The results can be understood easily in comparison with the true bilayer quantum Hall system. Similarly to the latter,\cite{Narasimhan1995} the stacked triangular crystal is favored for large distances between the effective layers (\emph{i.e.} large well widths). Indeed, when the effective layers are well separated the lattice tries above all to minimize the intra-layer repulsion, which favors triangular lattices. The residual inter-layer repulsion is then accomodated by the relative displacement of the two triangular lattices in each component. Conversely, when the effective layers get closer to each other, the lattice configuration tries to accomodate simultaneously the inter- and inter-layer interactions.
Therefore the overall lattice, which can be viewed as a bipartite lattice with one sublattice per layer, should be as closely packed as possible. This is not the case of the two interpenetrating triangular lattices that form a honeycomb lattice, and a stacked square lattice is therefore favored. Notice that an overall triangular lattice, which would have the closest packing, cannot be formed at finite separations between the effective layers because it is not bipartite. It might eventually be realized in the putative limit where the two effective layers are no longer spatially separated and where the intra- and interlayer repulsions are equal.

\subsection{Phase diagram of mono- and twocomponent Wigner crystals}

\begin{figure}[ht]
\begin{center}
\includegraphics[scale=1]{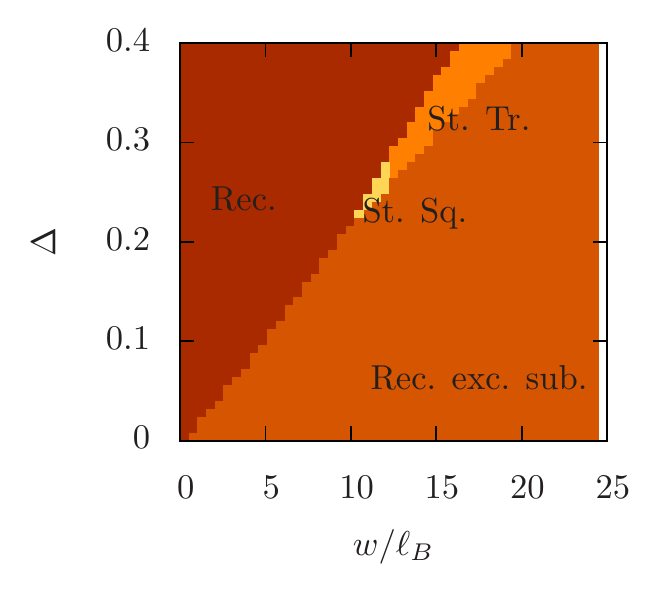}
\caption{(Color online) Wigner crystals phase diagram. The parameters are the width of the well $w$ and the subband gap $\Delta$ (here in units of $e^2/4\pi\epsilon\epsilon_0\ell_B$). ``Rec.'' denotes the rectangular WC in the lowest subband, ``Rec. exc. sub.'' is its \emph{alter ego} in the first excited subband, ``St. Tr.'' designs the stacked triangular WC and ``St. Sq.'' stands for stacked square.}
\label{fig:SolidPhaseDiagramWidthDelta}
\end{center}
\end{figure}

The WCs phase diagram presented in Fig.~\ref{fig:SolidPhaseDiagramWidthDelta} shares common features with the one of liquids (Fig.~\ref{fig:LiquidPhaseDiagramWidthDelta}). In large wells and for small subband gaps $\Delta$ the excited subband is occupied due to the reduced interaction energy, whereas in thin wells and for large subband gaps, a rectangular WC in the lowest subband has the lowest energy. Between these two limits of monocomponent states, one finds a region where stacked WCs can be stabilized (notably only in wells wider than $10\,\ell_B$), in the square version for $w/\ell_B \lesssim 13$, and in the triangular one in the top right orange region of the diagram. The stacked WCs are indeed favored by inter-component correlations and are thus reminiscent of the $(331)$ liquid state. As for the twocomponent states in the liquid phase diagram (Fig.~\ref{fig:LiquidPhaseDiagramWidthDelta}), the stacked crystals are encountered in a region that roughly obeys the scaling behavior shown in Eq. (\ref{Eq:scaling}).

\section{Phase diagram of the half-filled lowest Landau level -- Comparison between liquid and solid phases\label{sec:PD}}

\begin{figure}[ht]
\includegraphics[scale=1]{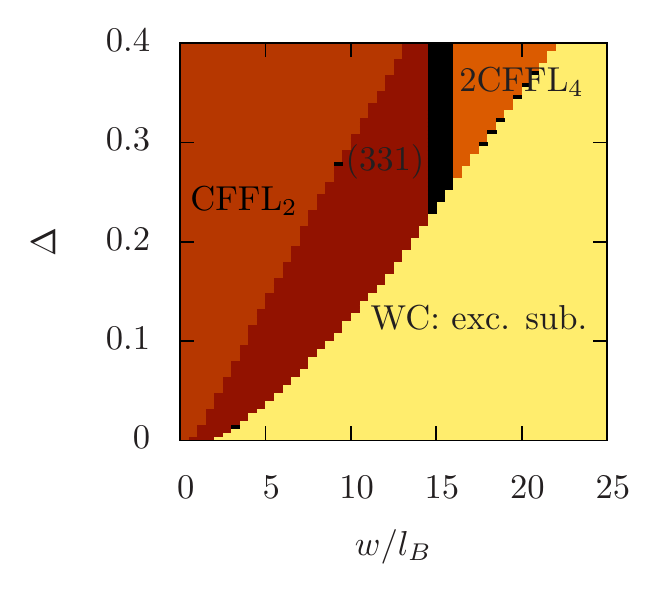}
\caption{(Color online) Phase diagram of the wide quantum well at $\nu=1/2$. Here the parameters are the width $w$ in units of the magnetic length $\ell_B$ and the subband gap $\Delta$ in units of the typical Coulomb energy $e^2/4\pi\epsilon\epsilon_0\ell_B$.}
\label{fig:PhaseDiagramWidthDelta}
\end{figure}

In the past two sections, we have presented an extensive set of model states for the 2DES at $\nu=1/2$. Whereas the energies of the various liquid states were computed by means of the VMC method in the spherical geometry, the energies of the solid states were obtained in the Hartree-Fock approximation in the planar geometry. We now give a detailed comparison of all investigated phases and draw a phase diagram for the competing liquid and solid states. The spherical and disc geometries are expected to lead to the same energies in the thermodynamic limit, which can thus be compared directly. The resulting phase diagram is shown in Fig.~\ref{fig:PhaseDiagramWidthDelta} as a function of the well width $w/\ell_B$ and the subband gap $\Delta$, which we continue to keep as an independent parameter for the moment. Whereas for narrow wells and large subband gaps, we retrieve, as expected from previous studies,\cite{Park1998,Papic2009,Biddle2013} liquid ground states in the half-filled lowest LL, these liquid phases disappear in favor of a rectangular WC in the first excited subband in the limit of large wells and small subband gaps. In addition to the usual CFFL$_2$ state in the lowest subband, the $(331)$ Halperin state remains stable in a part of the intermediate region, where both subbands are occupied even if it occupies a less prominent part of the phase diagram as compared to that (Fig. \ref{fig:LiquidPhaseDiagramWidthDelta}) where only liquid phases were considered. Note that the 2CFFL$_4$ state may also be stabilized in this intermediate region, albeit for $\Delta \ge 0.3$ and $w/\ell_B\ge 13$, i.e. for narrow wells and very large magnetic fields that are not realized experimentally.

In order to make a closer connection with experiments, we need to plot the phase diagram in Fig. \ref{fig:PhaseDiagramWidthDelta} in terms of experimentally relevant parameters. In contrast to our theoretical description, the subband gap in units of $e^2/4\pi\epsilon\epsilon_0\ell_B$ and the well width in units of the magnetic length $\ell_B$ cannot be varied independently. First, the well width is fixed by the sample-fabrication process, and one therefore needs to modify the ratio $w/\ell_B\propto \sqrt{B}$ by varying the magnetic field $B$, i.e. an increase of the magnetic field drives the system towards the large-well limit. Moreover, in order to keep the filling factor $\nu=n_{\mathrm{el}}/(eB/h)$ constant when varying $B$, one has to change the electronic density $n_{\mathrm{el}}$ simultaneously by applying an external gate voltage to the sample. An increase in $B$, however, also raises the energy unit $e^2/4\pi\epsilon\epsilon_0\ell_B\propto \sqrt{B}$, thus reducing effectively the subband gap $\Delta$ in our units. Hence for a given width $w$, increasing $B$ (and consequently $n_{\mathrm{el}}$) results in a larger width $w/\ell_B$ and a reduced subband gap $\Delta$ in our phase diagram (Fig. \ref{fig:PhaseDiagramWidthDelta}). 

\begin{figure}[ht]
\includegraphics[scale=1]{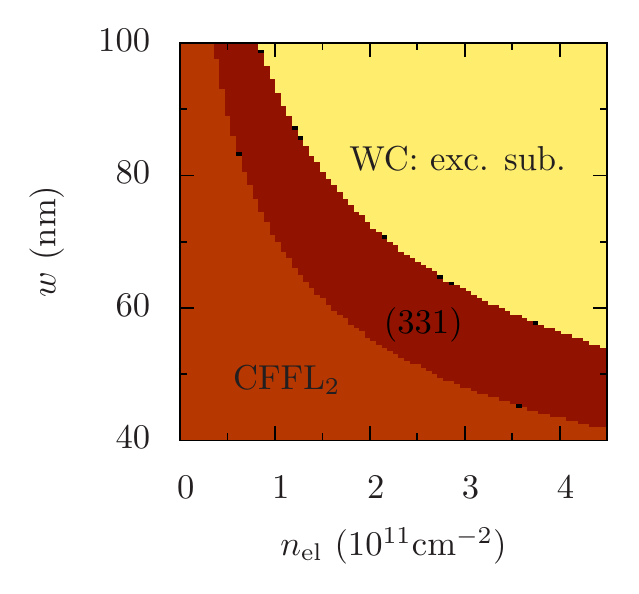}
\caption{(Color online) Theoretical phase diagram as a function of the experimentally relevant parameters, the width $w$ in nm and density $\nel$ in units of $10^{11}$ cm$^{-2}$.}
\label{fig:PhaseDiagram}
\end{figure}

\begin{figure}[ht]
\includegraphics[width=5cm]{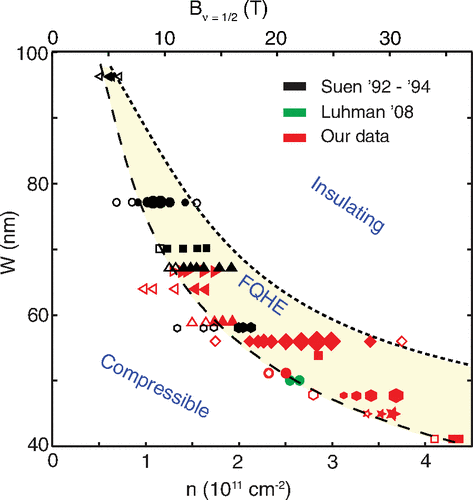}
\caption{(Color online) Experimental phase diagram of the wide quantum well at $\nu=1/2$, taken from Ref.~\onlinecite{Shabani2013}. Here $W$ in the width of the well, and $n$ is the electronic density.}
\label{fig:ExpPhaseDiagram}
\end{figure}

The phase diagram as a function of realistic values for the well width $w$ and the electronic density $\nel$ is plotted in Fig. \ref{fig:PhaseDiagram}, in comparison with a phase diagram by Shabani \textit{et al.} \cite{Shabani2013} for measured compressible, incompressible and insulating phases (Fig. \ref{fig:ExpPhaseDiagram}). As mentioned above, at a fixed value of the well width, the low-density limit corresponds to rather low magnetic fields where the subband gap $\Delta$ is substantial when compared to the interaction energy scale. The limit of large subband gaps and small well widths is therefore located in the lower left corner of the phase diagram, whereas the large-$w$/small-$\Delta$ limit is found in the upper right part. The scaling behavior (\ref{Eq:scaling}) of the intermediate region separating these to limits now translates to $w\propto 1/\nel\propto 1/B$ in the phase diagram in Fig. \ref{fig:PhaseDiagram}. Furthermore, as mentioned above, one notices that the only relevant twocomponent phase in this intermediate region is the incompressible $(331)$ Halperin state the 2CFFL$_4$ being shifted to parameters that are not investigated experimentally. Before comparing our findings in more detail with the experimental phase diagram (Fig. \ref{fig:ExpPhaseDiagram}), we point out that we have checked explicitly, within an enlarged model taking into account the third subband, that the latter is only populated for very large wells that are usually not studied experimentally ($w\gtrsim 100$ nm or $n_{\mathrm{el}}\gtrsim 4.5\times 10^{11}$ cm$^{-2}$), i.e. beyond the parameters for which we have presented the phase diagrams.

In experiments one does not have direct access to the nature of the ground state, but only to its transport characteristics. The experimental phase diagram in Fig. \ref{fig:ExpPhaseDiagram} from Ref. \onlinecite{Shabani2013} shows that a compressible state in the limit of narrow quantum wells and low electronic densities is separated from an insulating state at large $w$ and high densities via an intermediated region where an incompressible FQHE is observed. This is in excellent agreement with our phase diagram, where the FQHE is attributed to a $(331)$ state and the compressible one to a CFFL$_2$. Notice that even the lines separating the different regions are in good quantitative agreement. Whereas the compressible and incompressible states have commonly been understood in terms of a CFFL$_2$ and the $(331)$ state, respectively,\cite{Papic2009,Shabani2013} the origin of the insulating state remained unexplaned even if a WC seemed a natural candidate in view of the pertinence of correlations at half filling. Our results indicate that the measured insulating state can indeed be understood in terms of a crystalline state. Most saliently, it is a rectangular WC that is formed in the first excited electronic subband due to the confinement potential. Our results indicate that the effective interaction potential is modified by the subband wave functions in such a manner that crystalline rather than liquid phases are favored there, whereas crystalline phases at half filling are usually encountered in higher LLs in the form of stripe phases.\cite{Fogler1996,Fogler1997,MC,Lilly1999a,Lilly1999b,Du1999} Notice furthermore that the rectangular shape of the WC in the first excited subband is reminiscent of the stripe phase when considering the electronic density, especially since the latter in higher LLs reveals an unstability towards a density modulation along the stripes.\cite{Cote2000}

\section{Phase diagram of the half-filled second Landau level}\label{sec:LL1}

In this last section, we present the phase diagram for electron solids and quantum liquids for the half-filled second LL. From a technical point of view, the difference between the two LLs arises from the difference in the effective interaction potentials, which take into account the orbital wave functions associated with the LL. 
Concerning the VMC calculation of the liquid phases in the second LL, we use a different interaction than in the lowest one. Whereas, in the lowest LL, we could use a polynomial interaction (see the appendix) that reproduces exactly the pseudopotentials of the effective interaction (\ref{eq:potential}), we have no simple formula to compute the pseudopotentials in the second LL associated with monomials $r^k$ on the sphere. Thus, similarly to Ref.~\onlinecite{Park1998}, we use an alternative Gaussian interaction.
%\begin{equation}
%V_\mathrm{Gauss.} (r)= \frac{C_{-1}(w)}{r}+\sum_{k=0}^4 C_k(w) r^{2k} \er^{-r^2},
%\end{equation}
%that is reminiscent of that in Eq. (\ref{eq:PolynomialInt}), where the polynomial terms are multiplied by the Gaussian $\exp(-r^2)$. 
The coefficients are again chosen such that the associated pseudopotentials, now computed in the planar geometry, match the first five odd pseudopotentials of the effective interaction (\ref{eq:potential}).

\begin{figure}[ht]
\includegraphics[width=6cm]{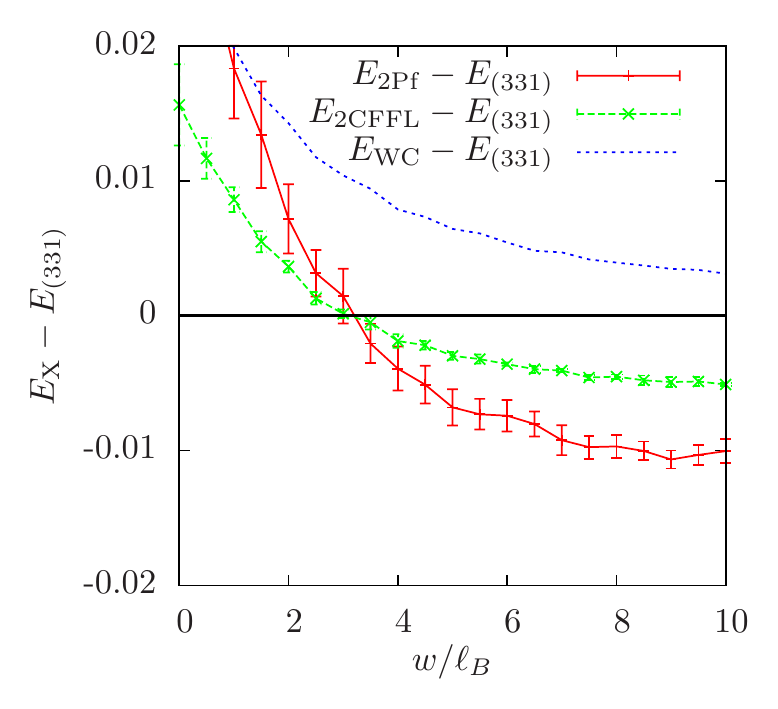}
\caption{(Color online) Energy differences of various intrinsic twocomponent states, as a function of the well width. We have chosen the $(331)$ state as the reference state (black continuous line).}
\label{fig:BilayerStates2LL}
\end{figure}

Figure \ref{fig:BilayerStates2LL} shows the energy differences for some of the states, with respect to the $(331)$ state. Here, we have chosen, for illustration reasons, to show only intrinsic twocomponent states that are insensitive to the subband gap. One notices that the stacked WC remains higher in energy than the different twocomponent FQHSs. Furthermore, we have checked that a phase of alternating stripes in the effective-bilayer configuration is even higher in energy than the stacked WC. Therefore in the intermediate region that separates the different monocomponent phases, similarly to the phases in the lowest LL, one finds a competition between the $(331)$ state, which is favored in narrow wells, and the 2Pf$_4$ in wide quantum wells, with a transition at $w/\ell_B\sim 3$. We emphasize that the incompressible 2Pf$_4$ state is a specificity of the second LL and is not stabilized in the lowest one.  

\begin{figure}[ht]
\includegraphics[width=6cm]{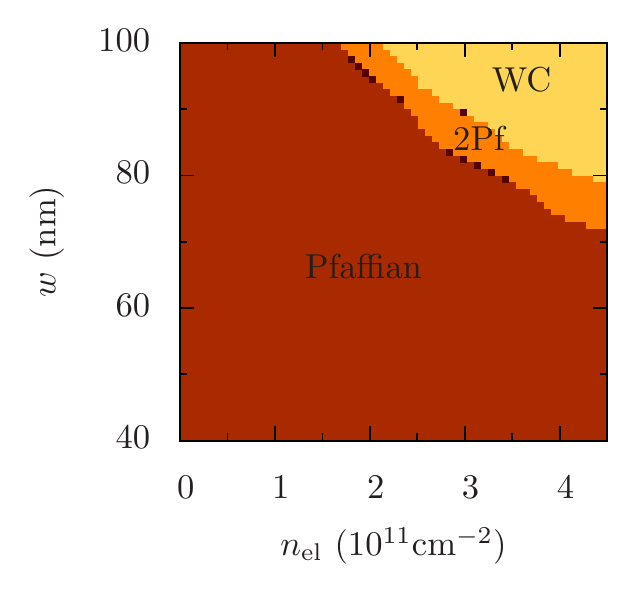}
\caption{(Color online) Phase diagram of the various states in the second LL as a function of the electronic density $\nel$ and the well width $w$. The (Pf$_2$) encountered phases are the monocomponent Pfaffian in the narrow-well limit, the rectangular WC in the limit of wide quantum wells, whereas an incompressible state (2Pf$_4$) of two four-flux Pfaffians (one in each subband component $+$ and $-$ is stabilized in an intermediate regime.) }
\label{fig:PhaseDiagram2LL}
\end{figure}

The phase diagram for the phases in the second LL is depicted in Fig. \ref{fig:PhaseDiagram2LL}, as a function of the experimentally relevant parameters $w$ and $\nel$. One notices that of the above-mentioned twocomponent states in the intermediate region of vanishing effective subband gap only the incompressible 2Pf$_4$ phase survives, whereas the $(331)$ state is covered by the usual monocomponent Pfaffian Pf$_2$ in the narrow-well limit, which extensive theoretical work\cite{Peterson2010a,Papic_Bilayer,Peterson2010b,Biddle2013,Morf1998} showed to be at the origin of the incompressible FQHS observed in this limit.\cite{Willett1987} 
Notice furthermore that the Pfaffian origin of the $5/2$ FQHS has recently been discussed intensely based on
effective interaction potentials modified by LL mixing that breaks the particle-hole symmetry of the half-filled
LL.\cite{Levin2007,Lee,Wang2009,Bishara2009,Wojs2010,Rezayi2011,Papic2012,Simon2013,Peterson2013,Sodemann2013}
As a consequence, it was argued that the anti-Pfaffian (the hole version of the Pfaffian) would be a more 
appropriate groundstate wavefunction at this filling factor.\cite{Levin2007,Lee,Wang2009,Bishara2009,Rezayi2011,Papic2012}
However, a possible transition between Pfaffian and anti-Pfaffian sensitively depends on the degree of LL mixing,
eventually altered by finite width effects, and for some of these parameters the Pfaffian is indeed 
stabilized.\cite{Wojs2010,Pakrouski2015}
While this debate is still ongoing, we emphasize that we neglect LL mixing within our model, such that the
monocomponent states are particle-hole-symmetric, and Pfaffian and anti-Pfaffian therefore 
have the same energy.\cite{zaletel2015}
Whereas the FQHSs are therefore different from those in the lowest LL, we obtain the same rectangular WC in the first excited subband as the ground state in the limit of wide quantum wells and large densities, in agreement with the insulating state in the lowest LL. However, we emphasize that this WC is encountered at larger values of $w$ and $\nel$ than in the lowest LL. % Finally we notice that the identification of the 2Pf$_4$ state is restricted to the variational approach we use here. We cannot exclude other possible states that we have not taken into account be lower in energy than all of those considered in our calculations. For a full identification of the incompressible state in the intermediate regime between the one-component Pfaffian and the WC in the second electronic subband, further calculations (e.g. exact diagonalization of density-matrix renormalization group) are required that are beyond the scope of the present paper.

\section{Conclusion}

In conclusion, we have compared the energies of various model states for the 2DES at $\nu=1/2$ in the lowest LL, taking into account the finite width of the confining quantum well explicitly within a square quantum well model. Keeping only the lowest two subbands, we obtain an effective twocomponent model of electrons restricted to a single LL in which the subband wave functions modify the effective interaction potential. This approximation is justified for the investigated parameter range, which covers that of available experimental
data and where we have checked explicitly that higher subbands beyond the second one are unoccupied. Within our twocomponent model, we have investigated in a variational approach the energies of various candidates for the ground state when varying the well width as well as the subband gap. This approach consists of VMC calculations for the energies of the different liquid states, such as the compressible CFFL, twocomponent CFFL (2CFFL$_4$) and the incompressible $(331)$-Halperin, Pfaffian and twocomponent Pfaffian (2Pf$_4$) states, that we compare to Hartree-Fock calculations for the energies of the electron solids, namely the various mono- and twocomponent WCs. 

Globally, one finds two limiting regions that can be seen as effective monocomponent limits. The first one is that of low well width and large subband gaps in which case all electrons reside in the lowest electronic subband. In this limit, which has largely been studied in the literature, one retrieves the compressible CFFL in the lowest LL. The second (monocomponent) limit is that of large well widths and low electronic subband gaps, where it is energetically favorable to populate the excited subband. The energy cost due to the subband gap is then overcome by the gain in interaction energy since the wave functions of this subband have a node that effectively reduces the Coulomb repulsion as compared to the lowest subband. Somewhat unexpectedly, one finds that crystalline phases are favored over the usual quantum liquids in this limit. Most saliently, we obtain a rectangular WC in the excited subband as the ground state in this limit. This phase is consistent with the experimentally observed insulating electronic behavior in this limit.\cite{Shabani2013}

The above-mentioned limiting regions of monocomponent states are separated by a relatively narrow region in which the subband gap is on the same order as the interaction-energy gain in populating the excited subband. In this region the equal population of both subband states, or more precisely of the symmetric and anti-symmetric bilayer-type superposition of both, allows for a reduction of the correlation energies. One finds then intrinsic twocomponent states, both if one considers crystalline phases, in the form of stacked crystals, and liquid phases, in the form of the $(331)$ Halperin state and a state consisting of two uncorrelated four-flux CFFL (2CFFL$_4$) in each of the two components. A comparison between these different phases shows that the liquid states are favored over the stacked crystals and that the incompressible $(331)$ state is realized in the region of experimentally accessible parameters. Our theoretical phase diagram is in excellent agreement with experimental findings for the different compressible, incompressible and insulating phases. First, our study confirms that one can understand the compressible phase in terms of the CFFL and the incompressible one as a $(331)$ state. Second it identifies a rectangular WC in the first excited subband as the origin of the insulating phase in the limit of large well width and low subband gap. From an experimental point of view, it is challenging to verify our image of a rectangular WC 
in the first excited subband as compared to other possible (crystalline) phases. In principle, a measure of the 
charge density in the $z$-direction would be required that is beyond the scope of present experimental techniques. 
Notice that recent microwave experiments have been interpreted in terms of a crystalline phase such as 
ours,\cite{Hatke2015} but an identification in terms of the particular crystal we obtain would require a detailed
study of the pinning mode, which might bear information about the charge profile in the $z$-direction. 
Furthermore, one may speculate that at intermediate temperatures the rectangular WC partially melts into a 
stripe-type phase restoring partially translation symmetry along the stripe direction. This would manifest itself in a highly anisotropic magnetoresistance as such measured in higher LLs.\cite{Lilly1999a,Lilly1999b,Du1999} 
However, these possible effects are mere speculation at this stage, and a detailed study of both the pinning mode and a possible partial melting is beyond the scope of the present paper.

Finally, we have investigated the competition between electron liquid and solid phases in the half-filled second LL. Similarly to the lowest one, monocomponent states in the narrow well are separated from monocomponent states in the wide-well small-subband-gap limit by an intermediate region of intrinsic twocomponent states. However, whereas one obtains the same insulating phase, which originates from a rectangular WC in the first excited subband, at large values of $w$ and $\nel$, the FQHSs are different from those in the lowest LL. Instead of a competition between a monocomponent CFFL and a twocomponent $(331)$ Halperin state, we encounter a competition between two Pfaffian phases in the second LL. In the narrow-well limit, we are confronted with the usual monocomponent Pfaffian state, which vanishes in the intermediate region in favor of the 2Pf$_4$ state that consists of two four-flux Pfaffian states in each effective-bilayer component. Whereas both liquid phases are incompressible, the WC in wide wells is expected to manifest itself via an insulating phase.

\acknowledgements

We would like to thank Csaba T\H oke for fruitful discussions, and Mansour Shayegan for his valuable comments and the permission to reproduce Fig.~\ref{fig:ExpPhaseDiagram}. N.R. was supported by a Keck grant and the Princeton Global Scholarship.

%-------------------------------------------------------------
\appendix
\section{Variational Monte-Carlo \label{sec:VMC}}

In this appendix we present the procedure we have used to determine the energies of the various liquid model states considered in this paper.

We compute the energies of the relevant FQHSs by direct integration of the trial wave functions in real space. For a system of $N$ electrons, the interaction energy in a model state $\Psi$ reads
\begin{multline}\label{eq:MCEnergy}
\L V \R_{\Psi} = \mathcal{N}\int d^2r_1 \dots d^2r_N \;V\left(\rv_1,\dots,\rv_N\right)\\ \times  \l|\Psi\left(\rv_1,\dots,\rv_N\right)\r|^2,
\end{multline}
where 
\begin{equation}\label{eq:MCNormalization}
\mathcal{N}=\left[ \int d^2r_1 \dots d^2 r_N \l|\Psi\left(\rv_1,\dots,\rv_N\right)\r|^2 \right]^{-1}
\end{equation}
is a normalization constant that is necessary since we deal with unnormalized wave functions. The integrals~(\ref{eq:MCEnergy}) and~(\ref{eq:MCNormalization}) are computed by VMC in the spherical geometry. 

Since we do not use a normal-ordered version of the interaction operator in Eq. (\ref{eq:IntOp}),
the interaction energy per particle diverges in the thermodynamic limit. We therefore have to introduce an explicit regularization energy that is common to all trial states in order to allow for a direct comparison between the latter. We arbitrarily set this regularization energy $E_\mathrm{reg.}$ to the homogeneous part of the interaction energy of an electron system that only populates the lowest subband
\begin{equation}\label{eq:RegularizationEnergy}
E_\mathrm{reg.}= \frac{1}{2} \int V^{1111}(r)\; d^2r.
\end{equation}
This energy also diverges in the thermodynamic limit, but if we substract it from the interaction energy~(\ref{eq:MCEnergy}) the interaction energy \emph{per particle} becomes finite. We call it the \emph{cohesive energy} $E_{\mathrm{coh.}}$, and will use this quantity in the following to compare the different phases, be they liquid or solid. We have checked that other choices for the reference energy do not change the energy differences between trial states.

\begin{figure}[ht]
\begin{center}
\includegraphics[scale=1]{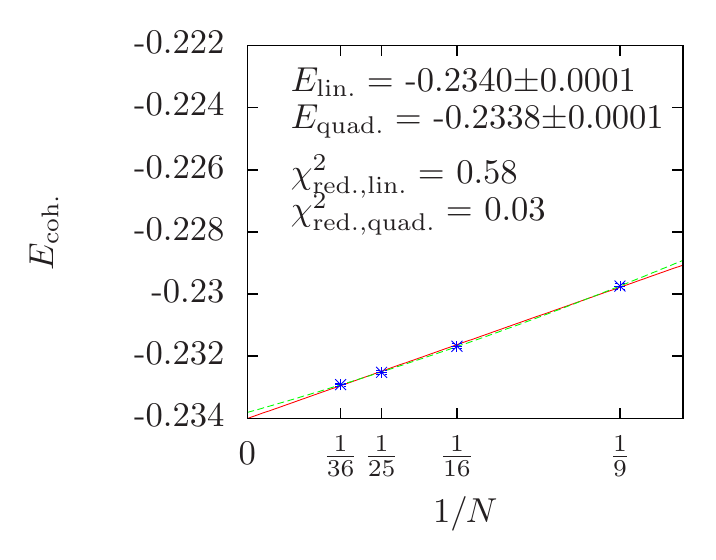}
\caption{(Color online) Example of extrapolation of the cohesive energy to the thermodynamic limit. Here the state is the CFFL in the lowest subband, and the width is $w= 10\,\ell_B$. Each blue point was obtained by averaging 10 VMC runs of $10^6$ iterations, hence the smallness of the error bars, which is of the size of the blue symbols. Since we consider only filled shells, we only access values of $N$ that are square of an integer. The solid red line is the result of the linear regression, and the dashed green line shows the second order polynomial regression. The cohesive energy is the value of the second-order polynomial for $1/N=0$, and the error is the difference of this value with the one given by the linear regression.}
\label{fig:ExempleInterpolation}
\end{center}
\end{figure}

For all accessible sizes we first compute the interaction energy~(\ref{eq:MCEnergy}) by VMC, then we substract the regularization energy~(\ref{eq:RegularizationEnergy}), which is finite for a finite size of the sphere. A division by the number of electrons $N$ gives the finite-size cohesive energy. Then we plot this quantity as a function of $1/N$ and perform a second-order polynomial regression. The cohesive energy in the thermodynamic limit is given by the value of this polynomial at $1/N=0$. The procedure is illustrated in Fig.~\ref{fig:ExempleInterpolation}, where a comparison with a linear regression to $1/N=0$ shows that the uncertainty ($<0.1$ \%) of the procedure is sufficiently small with respect to the energy difference of the trial states.

In the VMC method the interaction potential~(\ref{eq:potential}) is evaluated many times. In order to speed up computations we have used an alternative potential defined by
\begin{equation}\label{eq:PolynomialInt}
V_{\mathrm{poly.}}(r) = \frac{1}{R}\left[\frac{c_{-1}}{r/R} + \sum_{k=0}^{N_B} c_k \left(\frac{r}{R}\right)^k\right] ,
\end{equation}
where $N_B$ is the number of flux quanta and $R=\sqrt{N_B/2}$ is the radius of the sphere for the system size under consideration. Here, the coefficients $c_k$ are chosen to be such that the Haldane pseudopotentials of the polynomial interaction~(\ref{eq:PolynomialInt})
reproduce exactly the ones of the original interaction (see Ref.~\onlinecite{Papic2009} for details). Since the pseupotentials are identical, so are the interaction energies, such that this method does not change the computed energy values.

In order to analyze the twocomponents states, we have to determine two sets of coefficients $c_k$ for the polynomial interaction coefficients of Eq.~(\ref{eq:PolynomialInt}) defined above, one for the intra-layer interaction 
\begin{multline}\label{eq:IntraLayerEffPot}
V^{(\mathrm{A})}(r)=   \int_{-\tfrac{w}{2}}^{\tfrac{w}{2}}  \int_{-\tfrac{w}{2}}^{\tfrac{w}{2}} \dfrac{\l|\varphi_{+}(\zeta)\r|^2 \l| \varphi_{+}(\zeta')\r|^2}{\sqrt{r^2+(\zeta-\zeta')^2}} d\zeta d\zeta'
\end{multline}
between electrons that belong to the same effective layer [the effective intra-layer interaction is the same in both layers due to $\varphi_+(\zeta)=\varphi_-(-\zeta)$], and another set for the inter-layer interaction
\begin{multline}\label{eq:InterLayerEffPot}
V^{(\mathrm{E})}(r)=   \int_{-\tfrac{w}{2}}^{\tfrac{w}{2}}  \int_{-\tfrac{w}{2}}^{\tfrac{w}{2}} \dfrac{\l|\varphi_{+}(\zeta)\r|^2 \l| \varphi_{-}(\zeta')\r|^2}{\sqrt{r^2+(\zeta-\zeta')^2}} d\zeta d\zeta'.
\end{multline}
In principle one should also take into account the ``cross term''
\begin{multline}\label{eq:CrossLayerEffPot}
V^{(\mathrm{X})}(r)=   \int_{-\tfrac{w}{2}}^{\tfrac{w}{2}}  \int_{-\tfrac{w}{2}}^{\tfrac{w}{2}} \dfrac{\varphi_{+}(\zeta) \varphi_{-}(\zeta')\varphi_{-}(\zeta)\varphi_{+}(\zeta')}{\sqrt{r^2+(\zeta-\zeta')^2}} d\zeta d\zeta'.
\end{multline}
This term vanishes for a real bilayer but not for an effective bilayer formed in a quantum well, due to the non zero overlap of the $+$ and $-$ effective layer states.  
However, $V^{(\mathrm{X})}$ is more than an order of magnitude smaller than $V^{\mathrm{A}}$ or $V^{\mathrm{E}}$ (even for the largest well widths considered here), such that we neglect it in our variational calculation.

For completeness and a check of our numerical methods, we present in Tab.~\ref{tab:CoulombEnergies} our results for the liquid phases in the limit $w=0$ of the ideal 2DES in comparison with published results from Refs.~\onlinecite{Park1998} and \onlinecite{Biddle2013}. The results agree within the numerical error bars. 

\begin{table}[ht]
\centering
\begin{ruledtabular}
\begin{tabular}{cccc}
State & Ref.~\onlinecite{Park1998} & Ref.~\onlinecite{Biddle2013} & This work \\
\hline
%Laughlin & &\checkmark\\
 $(331)$ & ---  & $-0.4631(3)$ &  $-0.4633 \pm 0.001$ \\
Pfaffian & $-0.4569(2)$& $-0.4573(3)$ & $-0.4574\pm 0.003$ \\
CFFL &$-0.46557(6)$ & --- & $-0.4656 \pm 0.004$ 
\end{tabular}
\end{ruledtabular}
\caption{Cohesive energies of the liquid states we consider in the $w=0$ ``pure Coulomb'' limit. The first two columns were retrieved from existing publications, and the last column gives our results in this limit.}
\label{tab:CoulombEnergies}
\end{table}

%%%%%%%%%%%%%%%%

\section{Overlap corrections in the Wigner crystals}
\label{app:B}
\subsection{Preliminary: 2-body localized state}
A normalized coherent state centered around $\Rv$ reads 
\begin{equation}
\Psi_{\Rv}\l(\rv \r) = \l\L \rv|\Rv \r\R= \frac{1}{\sqrt{2\pi}} \er^{\frac{\ir}{2}\l(y X - x Y\r)} \er^{-\frac{\l(\rv -\Rv\r)^2}{4} }.
\end{equation}
Before writing the state for two electrons localized around $\vector{0}$ and $\Rv= R_0 \ev_x$, one should note that we have to antisymmetrize two one-body states that are not orthogonal ($\l\L \vector{0}| \vector{R}\r\R \neq 0$), hence if 
\begin{equation}
\l|\vector{0}, \vector{R}\r\R = \mathcal{N}\l(\l|\vector{0}\r\R\otimes \l|\vector{R}\r\R - \l|\vector{R}\r\R \otimes\l|\vector{0}\r\R\r)
\end{equation}
is the normalized and anti-symmetrized two-body state, 
\begin{equation}\begin{aligned}
\l\L \vector{0},\Rv | \vector{0},\Rv \r\R 
&= \mathcal{N}^2\l(2 \l \L \vector{0}|\vector{0} \r\R \l\L \vector{R}|\vector{R} \r\R -2 \l\L \vector{0}|\Rv \r\R\l\L\Rv| \vector{0} \r\R\r)\\
&= 2 \mathcal{N}^2\l(1-\l|\l\L \vector{0}|\Rv \r\R\r|^2 \r)
\end{aligned}
\end{equation}
the last term is not zero due to the non-orthogonality of coherent states, and it reads
\begin{equation}
\l\L \vector{0}|\Rv \r\R = \int \dr^2\rv  \;\Psi^*_{\vector{0}}\l(\rv \r)\Psi_{\vector{R}}\l(\rv \r) = \er^{-\frac{R^2}{4}}.
\end{equation}
Hence the normalization factor $\mathcal{N}$ is
\begin{equation}
\mathcal{N} = \l(1-\er^{-\frac{R^2}{2}}\r)^{-1/2},
\end{equation}
such that the two-body state centered around $\vector{0}$ and $\Rv$ finally reads
\begin{equation}
\begin{aligned}
\Psi_{2}\l(\rv_1,\rv_2 \r) &= \mathcal{N} \det
\begin{pmatrix}
\Psi_{\vector{0}}\l(\rv_1 \r) & \Psi_{\Rv}\l(\rv_1 \r)\\
\Psi_{\vector{0}}\l(\rv_2 \r) & \Psi_{\Rv}\l(\rv_2 \r)
\end{pmatrix}\\
&=\frac{\mathcal{N}}{2\pi}\l[ \er^{-\frac{\rv_1^2}{4}}\er^{\frac{\ir}{2}y_2 R_0 -\frac{\l(\rv_2 -\Rv\r)^2}{4}} \r. \\ & \l. \qquad \qquad- \er^{-\frac{\rv_2^2}{4}}\er^{\frac{\ir }{2}y_1R_0 - \frac{\l(\rv_1- \Rv \r)^2}{4}}\r].
\end{aligned}
\end{equation}
The probabilty density of this state is given by
\begin{equation}
\begin{aligned}
\rho\l(\rv\r)&= \l\L \psi^\dg\l(\rv\r)\psi\l(\rv\r)\r\R =  2 \int \l| \Psi_{2}\l(\rv , \rv_2 \r) \r|^2 \dr \rv_2 \\
%=  &\dfrac{1}{1-\er^{-\frac{R^2}{2}}} \frac{1}{\l(2\pi\r)^2} \int \l[ \er^{-\rv^2}{2}}\er^{-\frac{\l(\rv_2-\Rv\r)^2}{2}}+\er^{-\frac{\rv_2^2}{2}}\er^{-\frac{\l(\rv-\Rv\r)^2}{2}}- 2\cos \l( \frac{\l(y-y_2\r) R_0}{2} \r) \er^{-\frac{\rv^2 +\rv_2^2}{4}} \er^{-\frac{\l(\rv -\Rv\r)^2}{4}} \er^{-\frac{\l(\rv_2 -\Rv\r)^2}{4}} \r] \dr \rv_2 \\
= & \dfrac{\mathcal{N}^2}{2\pi} \bigg[ \er^{-\frac{\rv^2}{2}} + \er^{-\frac{\l(\rv-\Rv\r)^2}{2}}
- 2\cos \l( \frac{y R_0}{2} \r)\er^{-\frac{\rv^2}{4}}\er^{-\frac{\l(\rv -\Rv\r)^2}{4}} \\& \times\int \er^{-\frac{x_2^2+\l( x_2 -R_0 \r)^2}{4}}\dr x_2 \int  \cos \l( \frac{y_2R_0}{2} \r)\er^{-\frac{y_2^2}{2}}\dr y_2\bigg].
\end{aligned}
\end{equation}

This density can be written as the sum of two terms, $\rho(\rv)=\rho_0\l(\rv \r)  + \delta \rho\l(\rv\r)$, with
\begin{equation}
\rho_0\l(\rv \r) =\frac{\mathcal{N}^2}{2\pi}\l[ \er^{-\frac{\rv^2}{2}}+\er^{-\frac{\l(\rv-R_0\ev_x\r)^2}{2}} \r]
\end{equation}
and
\begin{equation}
\delta\rho \l(\rv \r) = -\frac{\mathcal{N}^2}{\pi} \cos \l( \frac{y R_0}{2} \r)\er^{-\frac{\l(\rv -\frac{R_0}{2}\ev_x\r)^2}{2}} \er^{-\frac{3 R_0^2}{8}}.
\end{equation}
In order to compute the interaction energy of this state in the Hartree-Fock approximation we express this density in the reciprocal space,
\begin{equation}
\begin{aligned}
\tilde{\rho}_0\l(\qv\r)& = \iint \rho_0\l(\rv \r) \er^{-\ir \rv\cdot\qv } \dr \rv \\
& = \mathcal{N}^2\er^{-\frac{\qv^2}{2}} \l[ 1 +\er^{\ir q_x R_0}\r] \\
& = 2\mathcal{N}^2\er^{-\frac{\ir q_x R_0}{2}}\er^{-\frac{\qv^2}{2}} \cos\l(\frac{q_x R_0}{2}\r)
\end{aligned}
\end{equation}
and
\begin{equation}
\begin{aligned}
\delta \tilde{\rho}\l(\qv\r) &= \iint \delta \rho\l(\rv \r) \er^{-\ir \l(xq_x+yq_y\r) } \dr x \dr y \\
&= -2\mathcal{N}^2\er^{-\frac{\ir}{2} q_x R_0}\er^{-\frac{R_0^2}{2}} \er^{-\frac{\qv^2}{2}} \cosh\l( \frac{q_y R_0}{2} \r) .
\end{aligned}
\end{equation}

\begin{figure}[ht]
\begin{center}
\caption{Density without corrections}
\includegraphics[scale=1]{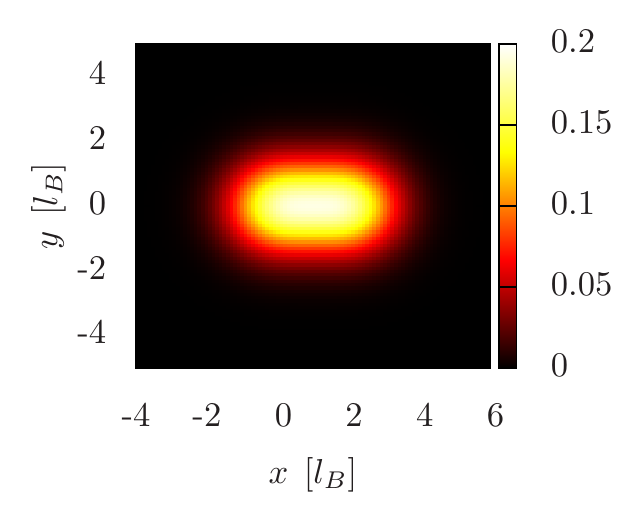}
\caption{Density with corrections}
\includegraphics[scale=1]{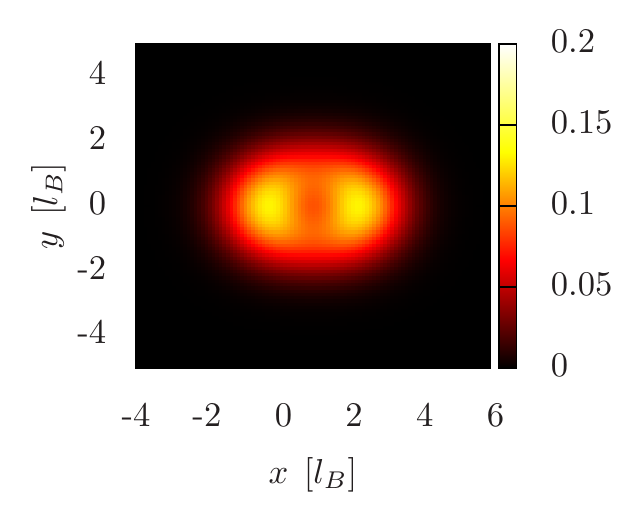}
\caption{(Color online) Corrections to the Wigner crystal density due to the overlap of neighbor states, localized around $\Rv_1=\vector{0}$ and $\Rv_2=2\ev_x$. The top figure shows the uncorrected density for comparison.}
\label{fig:DensiteCorrectionsWigner}
\end{center}
\end{figure}

This result can be generalized to a two-body state centered around any pair of locations $\Rv_1$ and $\Rv_2$,
\begin{equation}
\begin{aligned}
\Psi_{2}& \l(\rv_1,\rv_2 \r) = \frac{\mathcal{N}_2}{\sqrt{2}}\det
\begin{pmatrix}
\Psi_{\Rv_1}\l(\rv_1 \r) & \Psi_{\Rv_2}\l(\rv_1 \r)\\
\Psi_{\Rv_1}\l(\rv_2 \r) & \Psi_{\Rv_2}\l(\rv_2 \r)
\end{pmatrix} \\
&= \frac{\mathcal{N}_2}{\sqrt{2} 2\pi}\Big[ \er^{\frac{\ir}{2}\Rv_1\wedge \rv_1 -\frac{\l(\rv_1-\Rv_1\r)^2}{4}}\er^{\frac{\ir}{2}\Rv_2\wedge \rv_2-\frac{\l(\rv_2 -\Rv_2\r)^2}{4}} \\ &\qquad-\er^{\frac{\ir}{2}\Rv_2\wedge \rv_1 -\frac{\l(\rv_1-\Rv_2\r)^2}{4}}\er^{\frac{\ir}{2}\Rv_1\wedge \rv_2-\frac{\l(\rv_2 -\Rv_1\r)^2}{4}} \Big],
\end{aligned}
\end{equation}
where $\delta R = \l\| \Rv_1-\Rv_2 \r\|$ is the distance between the two centers and 
\begin{equation}
\mathcal{N}_2 = \l(1-\er^{-\frac{\delta R^2}{2}}\r)^{-1/2} .
\end{equation}
Finally
\begin{equation}
\begin{aligned}
\rho(\rv) & = \frac{\mathcal{N}_2^2}{2\pi} \Big[ \sum_{i=1,2}\er^{-\frac{\l(\rv-\Rv_i\r)^2}{2}}\\ &- 2 \cos \l(\frac{1}{2} \rv\wedge \delta \Rv \cdot \ev_z \r)\er^{-\frac{\l(\rv-\Rv\r)^2}{2}}\er^{-\frac{3\bm{\delta} \Rv^2}{8}}  \Big].
\end{aligned}
\end{equation}
where $\Rv=\l(\Rv_1+\Rv_2\r)/2$ et $\bm{\delta}\Rv=\Rv_1-\Rv_2$. This density if shown on Fig.~\ref{fig:DensiteCorrectionsWigner} for a two-body state localized around $\Rv_1=\vector{0}$ and $\Rv_2=2\ev_x$.
As one might have expected, the main effect of the overlap correction is to suppress the electronic density
between the two centers (i.e. at $\rv =\ev_x$), due to the antisymmetrization of the wave function.

The corresponding Fourier transform is given by $\tilde \rho = \tilde \rho_0 +\delta \tilde \rho$ where
\begin{equation}
\tilde \rho_0 (\qv) = \mathcal{N}_2^2 \er^{-\frac{q^2}{2}} \sum_{i=1,2} \er^{-\ir\qv\cdot\Rv_i} 
\end{equation}
and
\begin{equation}
\begin{aligned}
\delta \tilde \rho (\qv) & = -\frac{\mathcal{N}_2^2}{\pi}\er^{-\frac{3\delta R^2}{8}} \iint \dr \uv  \er^{-\ir\qv\cdot \l(\uv+\Rv\r)} \\
& \quad \times \cos\l( \frac{1}{2}\uv\wedge\bm{\delta}\Rv\r) \er^{-\frac{\uv^2}{2}}\\
&= -2\mathcal{N}_2^2 \er^{-\ir\qv\cdot \Rv}\er^{-\frac{\delta\Rv^2}{2}}\er^{-\frac{q^2}{2}} \cosh\l(\frac{1}{2}\qv\wedge\bm{\delta}\Rv\r) .
\end{aligned}
\end{equation}
This correction is thus proportional to $\er^{-\bm{\delta}\Rv^2/2}$ and generally negligible compared to the main term $\tilde{\rho}_0$, when the two centers $\Rv_1$ et $\Rv_2$ are separated by several magnetic lengths, because 
\[
\delta \tilde \rho (\qv) / \tilde \rho_0 (\qv) \propto \er^{-\bm{\delta}\Rv^2/2}.
\]

\subsection{Many-body generalization}
The $N$-electrons Wigner crystal wave function is given by the Slater determinant of coherent states
\[
\begin{aligned}
\Psi_{N}\l(\rv_1,\rv_2,\dots,\rv_N \r) &= \mathcal{N}_N\det
\l(
\Psi_{\vector{R}_i}\l(\rv_j \r) \r)
 \\&=\mathcal{N}_N \sum_{p\in \sigma_N} \mathrm{sgn}(p) \prod_{i=1}^N \Psi_{\Rv_i}\l(\rv_{p(i)}\r),\end{aligned}
 \]
where $\Rv_i$ are the $N$ sites of the underlying lattice, and $\mathcal{N}_N$ is a normalization factor.
The corresponding density is
\begin{equation}\label{eq:DensiteGeneral}
\begin{aligned}
\rho_N&\l(\rv_1\r)= N!	\int \dots \int
\l|\Psi_{N}\l(\rv_1,\rv_2,\dots,\rv_N \r)\r|^2 \dr\rv_2 \dots \dr\rv_N\\
&\quad=\mathcal{N}_N\times N!\times\sum_{p,p'\in \sigma_N} \mathrm{sgn}(p)\times  \mathrm{sgn}\l(p'\r) \\
&  \times\int\dots \int \prod_{i,j=1}^N \Psi_{\Rv_i}^*\l(\rv_{p(i)}\r) \Psi_{\Rv_j}\l(\rv_{p'(j)}\r) \dr\rv_2 \dots \dr\rv_N.
\end{aligned}
\end{equation}
At this step it is very common to suppose that the following terms
 \[
 \int \Psi_{\Rv_\alpha}^*\l(\rv_i\r) \Psi_{\Rv_\beta}\l(\rv_i\r) \dr\rv_i \propto \delta_{\alpha,\beta}
 \]
are negligible if $ \alpha \neq \beta$, \emph{i.e.} the overlap of neighboring localized states $\Rv_\alpha$ and $\Rv_\beta$ is neglected because their probability density decreases exponentially with distance. Within this approximation, which is justified in the low-density limit ($\nu\ll 1$), the density~(\ref{eq:DensiteGeneral}) is simply given by
\begin{equation}
\begin{aligned}
\rho_N\l(\rv_1\r) &= \sum_{p\in\sigma_N} \prod_{i=1}^N \int\dots\int \l|\Psi_{\Rv_i}\l(\rv_{p(i)}\r)\r|^2 \dr\rv_2 \dots \dr\rv_N \\
&= \sum_{i=1}^N \l|\Psi_{\Rv_i}\l(\rv\r)\r|^2.
\end{aligned}
\end{equation}
This density is the one of a direct product state
\[\Psi_{\mathrm{prod.}}\l(\rv_1,\dots,\rv_N\r)=\Psi_{\Rv_1}\l(\rv_1\r)\Psi_{\Rv_2}\l(\rv_2\r)\dots\Psi_{\Rv_N}\l(\rv_N\r),\]
which has a large overlap with the real fermionic state~(\ref{eq:DensiteGeneral}) although it is not antisymmetric.

For large densities, $\nu\simeq 1/2$, the approximation to neglect the overlap of neighboring states is no longer valid. This is particularly true for the anisotropic crystals such as the rectangular and rhombic Wigner crystals, where neighboring sites may be very close to each other ($\delta R \simeq 2\,\ell_B$). 
While it would be tedious to take into account all the terms in the density~(\ref{eq:DensiteGeneral}), we can at least take into account the overlap between nearest neighbors. 
Explicitely, under this approximation we suppose the terms
\[
 \int \Psi_{\Rv_\alpha}^*\l(\rv_i\r) \Psi_{\Rv_\beta}\l(\rv_i\r) \dr\rv_i
 \]
are non-zero if $\alpha=\beta$, as usual, but also if $\Rv_\alpha$ and $\Rv_\beta$ are nearest-neighbor sites of the crystal lattice. The integrals involved in the calculation of the latter terms can be deduced from the two-body calculations presented in the previous section. 

The Fourier transform of the many-body state reads
\begin{equation}
\begin{aligned}
\tilde{\rho}(\qv)& = \frac{1}{1-2\er^{-\frac{\delta R^2}{2}}}\er^{-\frac{q^2}{2}} \Bigg[ 
%\sum_i \er^{-\ir\qv\cdot\Rv_i} 
\sum_j \delta\l(\qv-\Qv_j\r)\\
&-2\sum_{\langle m,n\rangle}\er^{-\frac{\bm{\delta}\Rv^2}{2}} \er^{-\ir\qv\cdot \Rv}\cosh\l(\frac{1}{2}\qv\wedge\bm{\delta}\Rv\r) \Bigg],
\end{aligned}
\end{equation}
where $\langle m,n \rangle$ refers to first neighboring sites, with relative coordinates $\bm{\delta}\Rv = \Rv_m-\Rv_n$ and center of mass $\Rv=\l(\Rv_m+\Rv_n\r)/2$, where $\Qv_j$ are the vectors of the reciprocal lattice.  
The normalization term $[1-2\exp(-\delta R^2/2)]^{-1}$ was determined \emph{a posteriori} with the constraint 
\begin{equation}
\iint \rho\l(\rv\r) \dr\rv = \tilde{\rho}(\qv=\vector{0})= N,
\end{equation}
and we have used $\sum_i \er^{-\ir\qv\cdot\Rv_i}=(2\pi)^2\sum_j \delta\l(\qv-\Qv_j\r)/A_{u.c.}$,
where $A_{u.c.}$ is the unit-cell area of the Wigner crystal.
%\begin{figure}[ht]
%  \begin{center}
%  \includegraphics[width=0.2\textwidth]{Fig17A.pdf}
%    \quad \includegraphics[width=0.2\textwidth]{Fig17B.pdf}
%  \caption{Elementary cell of the rectangular and rhombic lattices.\label{fig:ReseauRectangulaireAnnexe}}
%  \end{center}
%\end{figure}

%We first compute the corrections for the rectangular and rhombic Wigner crystals, since for the other particular lattices the nearest neighbors are sufficiently distant from one another to neglect these corrections.
%\[
%\begin{aligned}
%\delta R &= \sqrt{2\pi/\nu} \le \sqrt{4\pi} \simeq 3,54 \;\ell_B \\
%&  \qquad \Longrightarrow \qquad \frac{\delta \tilde \rho (\qv)}{ \tilde \rho_0 (\qv)}  \propto \er^{-\delta R^2/2} \le 0,002
%\end{aligned}
%\]
%and for the triangular lattice this correction is even smaller because the first neighbors are even farther from each other ($\delta R = \sqrt{4\pi/\sqrt{3}\nu}$).

We only consider a rectangular Wigner crystal that we have identified to be the crystal symmetry of lowest energy.
In a first step, we consider the pairs along the $y$ direction, for which the lattice spacing $\bm{\delta}\Rv = \delta Y \ev_y= R_0\ev_y/\sqrt{\alpha}$ is decreased, with our choice $\alpha>1$ for the lattice anisotropy parameter. (Similarly, we have an enhanced lattice spacing $\delta X = \sqrt{\alpha}R_0$ in the $x$-direction.)
Hence if we write the 
\begin{equation}
\Rv_{jk} =  j\delta X \ev_x + k\delta Y \ev_y   \qquad \{j,k\}\in \mathbb{Z}^2,
%\Rv_{jk} =  j\delta X \ev_x + \l(k+\frac{1}{2}\r)\delta Y \ev_y   \qquad \{j,k\}\in \mathbb{Z}^2,
\end{equation}
we obtain
\begin{equation}
\begin{aligned}
\delta \tilde{\rho}\l(\qv\r) &= -\dfrac{2\er^{-\frac{q^2}{2}}}{1-2\er^{-\frac{\delta R^2}{2}}} \er^{-\frac{\delta Y^2}{2}}\cosh\l(\frac{q_x \delta Y}{2}\r)\sum_{j,k}\er^{-\ir\qv\cdot \Rv} \\
&= -\frac{2}{1-2\er^{-\frac{\delta R^2}{2}}}\er^{-\frac{q^2}{2}} \er^{-\frac{\delta Y^2}{2}}\cosh\l(\frac{q_x \delta Y}{2}\r) \\
& \times \sum_{Q_x} \delta\l(q_x-Q_x\r) %\er^{-\ir q_y \delta Y/2}
\sum_{Q_y} \delta\l(q_y-Q_y\r)
\end{aligned}
\end{equation}
where $Q_x$ and $Q_y$ are defined by
\begin{equation} Q_x = \frac{2\pi}{\delta X} n =\frac{2\pi}{\sqrt{\alpha} R_0}n 
\end{equation}
and
\begin{equation}
Q_y= \frac{2\pi}{\delta Y} m =\frac{2\pi\sqrt{\alpha}}{R_0} m .
\end{equation}

%%%%%%%%%%%%%%%
\iffalse
Finally the corrections in the reciprocal space is given by
\begin{equation}
\begin{aligned}
\delta \tilde{\rho}\l(\qv\r)=&-\dfrac{2}{1-2\er^{-\frac{\delta Y^2}{2}}}\er^{-\frac{q^2}{2}} \er^{-\ir q_y \delta Y/2}\er^{-\frac{\delta Y^2}{2}} \\
&  \times \cosh\l(\frac{\delta Y}{2} q_x\r)\sum_{\Qv} \delta\l(\qv-\Qv\r),
\end{aligned}
\end{equation}
and the term 
\begin{equation}
\l| \l\L\bar{\rho}\l(\qv\r) \r\R \r|^2=
 \l| \l\L\tilde{\rho}\l(\qv\r) \r\R \r|^2\er^{q^2/2}=
\er^{-q^2/2}\sum_{\Qv} \delta\l(\qv-\Qv\r) 
\end{equation}
that appears in the WC energy expressions becomes
\begin{equation}
\begin{aligned}
\l| \l\L\bar{\rho}\l(\qv\r) \r\R \r|^2 =& \dfrac{\er^{-q^2/2}}{\l(1-2\er^{-\frac{\delta R^2}{2}}\r)^2} \sum_{\Qv}\delta\l(\qv-\Qv\r) \\ &
\times \Big[ 1 - 4 \cos\l( \frac{\delta Y}{2}q_y \r)\er^{-\frac{\delta Y^2}{2}}\cosh\l(\frac{\delta Y}{2} q_x\r) \\
 & \qquad +4  \er^{-\delta Y^2} \cosh^2\l(\frac{\delta Y}{2} q_x\r)\Big] .
\end{aligned}
\end{equation}

\fi
%%%%%%%%%%%%%%

If we also take into account the site paired in the $x$ direction, the correction to the projected density is
\begin{equation}
\begin{aligned}
\delta \tilde{\rho}\l(\qv\r)&=-\er^{-\frac{q^2}{2}}\sum_{\Qv}  \delta\l(\qv-\Qv\r)\\
&  \times \Big[ \dfrac{2}{1-2\er^{-\frac{\delta Y^2}{2}}} % \er^{-\ir q_y \delta Y/2}
\er^{-\frac{\delta Y^2}{2}} \cosh\l(\frac{\delta Y}{2} q_x\r)\\
&+\dfrac{2}{1-2\er^{-\frac{\delta X^2}{2}}} % \er^{-\ir q_x \delta X/2}
\er^{-\frac{\delta X^2}{2}} \cosh\l(\frac{\delta X}{2} q_y\r)\Big],
\end{aligned}
\end{equation}
and the squared density 
\begin{equation}
\l| \l\L\bar{\rho}\l(\qv\r) \r\R \r|^2=
 \l| \l\L\tilde{\rho}\l(\qv\r) \r\R \r|^2\er^{q^2/2}=
\er^{-q^2/2}\sum_{\Qv} \delta\l(\qv-\Qv\r) 
\end{equation}
appearing in the energy thus becomes

\begin{equation}\label{eq:density_corrections}
\begin{aligned}
&\l| \L \bar{\rho}(\qv) \R \r|^2 =  \l[1-2\l(\er^{-\frac{\delta X^2}{2}}+\er^{-\frac{\delta Y^2}{2}}\r)\r]^{-2} \sum_{\Qv}\delta\l(\qv-\Qv\r)\er^{-q^2/2}
\\ 
&\times \Bigg\{ 1 - 4\l[\er^{-\frac{\delta Y^2}{2}} % \cos\l( \frac{\delta Y}{2}q_y \r) 
\cosh\l(\frac{\delta Y}{2} q_x\r) 
%\r.\\ &\l.
+\er^{-\frac{\delta X^2}{2}} %\cos\l( \frac{\delta X}{2}q_x \r) 
\cosh\l(\frac{\delta X}{2} q_y\r)\r]  
\\
&+4   \l[\er^{-\delta Y^2}\cosh^2\l(\frac{\delta Y}{2} q_x\r)+\er^{-\delta X^2}\cosh^2\l(\frac{\delta X}{2} q_y\r)\r]
\\ 
&+8\er^{-\frac{\delta R^2}{2}}\l[ %\cos\l( \frac{\qv\wedge\bm{\delta} \Rv}{2}\r)
\cosh\l(\frac{\delta Y}{2} q_x\r) \cosh\l(\frac{\delta X}{2} q_y\r)\r] \Bigg\} .
\end{aligned}
\end{equation}

%where $\bm{\delta} \Rv = (\delta X, \delta Y)$ and  $\qv\wedge\bm{\delta} \Rv = q_x \delta Y - q_y \delta X$.

The expression~(\ref{eq:density_corrections}) takes into account first-order corrections to the non-overlapping approximation (that is usually made when computing WCs energies). Globally, these corrections energetically penalize strong anisotropies and/or large density, and we have used the expression~(\ref{eq:density_corrections}) for all our calculations.

\bibliography{these}

\end{document}